\documentclass[useAMS,referee,usenatbib]{biom}
\def\bSig\mathbf{\Sigma}

\RequirePackage[OT1]{fontenc}
\RequirePackage{natbib}
\RequirePackage[colorlinks,citecolor=blue,urlcolor=blue]{hyperref}
\usepackage{amsmath}
\usepackage{amsfonts}
\usepackage{graphicx,psfrag,epsf}
\usepackage{enumerate}
\usepackage{url} 
\usepackage{comment}

\newtheorem{proposition}[theorem]{Proposition} 
\usepackage{epsf,subfigure,pstricks,pst-node}
\usepackage{float}
\usepackage[outdir=./]{epstopdf}
\usepackage{verbatim} 

\def\BIC{\textsc{bic}}
\def\T{{ \mathrm{\scriptscriptstyle T} }}

\usepackage{times}
\usepackage{algorithm}

\title[Nonparametric Regression for Brain Imaging Data Analysis]{Nonparametric Matrix Response Regression with Application to Brain Imaging Data Analysis}

\author{Wei Hu$^1$, Tianyu Pan$^1$, Dehan Kong$^2$,  and Weining Shen$^{\ast,1}$\email{weinings@uci.edu}  \\ 
\textit{$^{1}$Department of Statistics, University of California, Irvine, California, U.S.A.}
\\ 
\textit{$^{2}$Department of Statistical Sciences, 
University of Toronto, Canada}
}

\begin{document}
\graphicspath{{./art/}}

\date{{\it Received October} 2007. {\it Revised February} 2008.  {\it
Accepted March} 2008.}



\pagerange{\pageref{firstpage}--\pageref{lastpage}} 
\volume{64}
\pubyear{2008}
\artmonth{December}


\doi{10.1111/j.1541-0420.2005.00454.x}


\label{firstpage}


\begin{abstract}
With the rapid growth of neuroimaging technologies, a great effort has been dedicated recently to investigate the dynamic changes in brain activity. Examples include time course calcium imaging and dynamic brain functional connectivity. In this paper, we propose a novel nonparametric matrix response regression model to characterize the nonlinear association between 2D image outcomes and predictors such as time and patient information. Our estimation procedure can be formulated as a nuclear norm regularization problem, which can capture the underlying low-rank structure of the dynamic 2D images. We present a computationally efficient algorithm, derive the asymptotic theory and show that the method outperforms other existing approaches in simulations. We then apply the proposed method to a calcium imaging study for estimating the change of fluorescent intensities of neurons, and an electroencephalography study for a comparison in the dynamic connectivity covariance matrices between alcoholic and control individuals. For both studies, the method leads to a substantial improvement in prediction error. \\
\end{abstract}

%
\vspace{5mm}


\maketitle


%

\section{Introduction}
\subsection{Motivation}
Neuroimaging techniques have been widely used in both scientific research and clinical applications, e.g., calcium imaging, electroencephalography (EEG), magnetic resonance imaging and functional magnetic resonance imaging. Those techniques provide a great opportunity for (1) the investigation of neuron activities at the cellular level, e.g., the quantitative estimation of the intracellular calcium signals; and (2) a better understanding of the dynamic changes in brain activity as well as their associations with disease and other clinical information. Here we briefly discuss two relevant examples. More details about these examples will be given in Section \ref{sec:dataapp}. The first example is related to the fluorescent calcium imaging, which is an emerging popular technique for observing the spiking activity of large neuronal populations over the past decade. This new technology allows scientists to observe the locations of neurons and the times at which they fire using two-photon microscopy \citep{helmchen2005deep, petersen2018scalpel}, resulting in a data set as a video clip, i.e., a sequence of 2D images taken over the time. The scientific questions of interest are to identify the major neurons and model their spiking activity over the time; and calcium imaging techniques provide a unique opportunity to address those questions because of its ability to directly collect the rich and accurate measurements at the cellular level.  As shown in Figure \ref{fig: bunchframe_scatter}(a), the images are quite noisy and structured-sparse (in the sense that the signal level is low at the boundary for most images), yet contains rich information (the video clip we study in this paper composes 3,000 picture frames with several obvious major neurons located in the center of the image firing for multiple times). Thus it is important and non-trivial to develop an automatic-yet-flexible pipeline for analyzing such type of data sets. 

Our second example is the brain functional connectivity analysis. Functional connectivity refers to the coherence of the activities among distinct brain regions \citep*{Horwitz2003}, and it provides novel insights on how distributed brain regions are functional integrated \citep{Biswal1995, Biswal2010, Fox2005}. In general, studies of functional connectivity are based on the temporal correlation between spatially remote neurophysiological events \citep{Friston1994} with an implicit assumption that the functional connectivity is at the same level during the observation period. Recently, functional connectivity has been shown to fluctuate over time \citep*{Chang2013}, implying that measures assuming stationarity over a full scan may be too simplistic to capture the full brain activity. Since the initial findings, researchers have investigated the so-called \emph{dynamic functional connectivity}; see \citep{Calhoun2014, Calhoun2016, Preti2016, zhang2016linear} for reviews to date. It then makes sense to represent the connectivity as a covariance matrix and model its change over the time and other clinical factors of interest. As shown in Figure \ref{fig:eeg_matrix}, we analyze an EEG data set to study the dynamic functional connectivity between alcoholic and non-alcoholic individuals. There is a significant difference in terms of image pattern and temporal correlation between two groups of participants.

\subsection{Challenges and contributions}
Statistically speaking, it is natural to quantify the association between 2D image outcomes and the predictors such as time, patient demographics, and other disease predictors by a regression model. Such regression model will also serve as a useful tool for smoothing and denoising purposes. However, several challenges remain unaddressed in the literature. First, the neuroimaging data obtained from the experiments, even after some pre-processing steps, are often noisy and taking complex structure, such as spatial/temporal correlation, low rankness, and sparsity. Such structural information provides useful scientific and clinical insight, but also imposes additional challenges to statistical analysis and computation. 



Secondly, existing regression approaches are mainly focused on linear models and image predictors in the literature, where the linearity assumption, as much as it simplifies the computational/theoretical investigation, does not quite meet the need to accurately model the nonlinear pattern between predictors and the image outcomes. For example, we performed a preliminary analysis on the calcium imaging data collected by Ilana Witten's lab at the Princeton Neuroscience Institute \citep{petersen2018scalpel}. Figure \ref{fig: bunchframe_scatter}(b) shows a scatter plot of the changes of fluorescent intensities across time from a randomly selected pixel of the 2D-image. The scatter plot shows a clear nonlinear pattern, which will be neglected by linear models. Note that classical nonlinear regression approaches such as Nadaraya-Watson method can not be directly used in our case to model the matrix-valued image responses, since doing so is equivalent to vectorizing the 2D-image data, which destroys the underlying spatial information of the image. 

In this paper, we propose a novel low rank nonparametric regression approach to take account of the matrix structure of the image data by solving a nuclear norm regularization problem. By the singular value thresholding algorithm \citep{cai2010singular}, we show that our estimator has a closed-form solution for each fixed bandwidth and regularization parameter. To efficiently select these tuning parameters, we derive a Bayesian information criterion (BIC) based on our model and estimation procedure. For theoretical justification, we derive the risk bound for our nonparametric estimator. We show that the rank of the true function can be consistently estimated as well.

Here we highlight our contributions in this paper. First, we propose a novel nonparametric matrix response regression model that is capable of modeling the nonlinear relationship between image responses and predictors. There are some related works on (generalized) linear models for matrix-valued data. For example, \citet{zhou2014regularized} proposed a class of regularized matrix linear regression model by treating matrix data as covariates; \citet{hu2020matrix} proposed a linear discriminant analysis approach for the classification of high-dimensional matrix-valued data; \citet*{wang2017} developed a generalized scalar-on-image regression model via total variation; \citet*{ding2018matrix} studied the matrix response linear regression model using envelope methods; \citet{kong2018l2rm} proposed a low-rank linear regression model with high-dimensional matrix response and high dimensional scalar covariates. \citet{li2017parsimonious,zhang2017tensor} studied the tensor linear regression problem with a multidimensional array response and a vector predictor. To the best of our knowledge, no work has been done on using nonparametric models for matrix data analysis. Second, our nonparametric estimator is easy to derive and has a closed-form solution, which makes it computationally more efficient than the state-of-art multivariate varying coefficient model \citep{zhu2011fadtts, zhu2012multivariate}. The tuning process is also simple thanks to the derived analytic form of BIC, which is not straightforward for the proposed nonparametric matrix response model. Thirdly, we develop the asymptotic theories including the risk bound and rank consistency for the method, which directly connects to the existing work on nonparametric statistics theory, hence rendering for a deeper understanding of such type of models. Finally,  as demonstrated in both real data analysis examples, the proposed nonparametric model is simple, ready-to-use, manages to fit the nonlinear changes in brain signals very well and has a substantial improvement in the prediction error. The low-rankness assumption seems amenable in practice. We are currently developing a computational package for the proposed methodology and we expect it to have a major impact in the neuroimaging community. 


\section{Data applications}\label{sec:dataapp}
In this section we provide more details about the two motivating data examples and highlight their scientific merit and associated statistical challenges. 
\subsection{Fluorescent calcium imaging}
Calcium ions generate versatile intracellular signals that control key functions in all types of neurons, including the control of
heart muscle cell contraction and the regulation of vital aspects of the entire cell cycle. Fluorescent calcium imaging is a sensitive method for monitoring neuronal activity \citep{mao2001dynamics}. It makes use of the fact that
in living cells, most depolarizing electrical signals are associated
with calcium ions influx attributable to the activation of one or more of
the numerous types of voltage-gated calcium ions channels, abundantly
expressed in the nervous system \citep{tsien1990calcium, stosiek2003vivo}. Imaging calcium in neurons is particularly important because calcium signals exert their highly specific functions in well-defined cellular subcompartments  \citep{grienberger2012imaging}. The advantage of calcium imaging is that it allows real-time analyses of individual cells and even subcellular compartments. At the same time it readily permits simultaneous recordings from many individual cells \citep{stosiek2003vivo}.

When a neuron fires, voltage-gated calcium channels in the axon terminal open and then calcium floods the cell. Such changes in concentration of calcium ions are detected by observing the fluorescence of calcium indicator molecules. Therefore, not surprisingly, intracellular calcium concentration becomes an important surrogate marker for the spiking activity of neurons in  the absence of  effective voltage imaging approach and is commonly used when analyzing local neuronal circuits in vivo and in vitro \citep{petersen2018scalpel, grienberger2012imaging}. However, the available experimental techniques still lead to noisy and spatial-temporally-subsampled observations of true underlying signals \citep{pnevmatikakis2012fast}. The calcium trace is smeared which restricts the extraction of quantities of interest such as spike trains of individual neurons \citep{rahmati2016inferring}. The spatio-temporal smoothing of the calcium profile remains a difficult problem due to the high dimensionality and complex structure of calcium imaging data. New methods that are capable of characterizing the dynamic changes of calcium imaging videos and performing the spatio-temporal smoothing of the calcium data are largely needed. 

\subsection{Dynamic functional connectivity}
Functional connectivity measures statistical dependence between the time series signals obtained from different brain regions. The functional connectivity has been investigated in a wealth of literature with various analysis tools, and has applications in different imaging modalities such as functional magnetic resonance imaging (fMRI), Electroencephalography (EEG), Magnetoencephalography (MEG) and Positron emission tomography (PET). The functional connectivity is built upon the assumption that the connectivity is stationary in nature. 

More recently, the dynamic behavior of functional connectivity was revealed, suggesting that connectivity between different brain regions exhibits meaningful variation over time. Hence it provides a more detailed representation of brain function than the traditional static functional connectivity analysis. Several previous studies have suggested that the dynamic functional connectivity can be used to characterize different psychiatric and neurodegenerative disorders such as schizophrenia \citep{damaraju2014dynamic}, Parkinson's disease \citep{engels2018dynamic, zhu2019abnormal} and Alzheimer's disease \citep{niu2019abnormal}. One of the most common ways to quantify the dynamic functional connectivity is by estimating the dynamic covariance matrices using a slide window approach. The dynamic covariance matrices have proven to be very useful in revealing the change patterns of brain network over the time. In addition, studying the association between those matrices with clinical factors and disease outcomes may provide useful insights for future clinical practice such as disease prediction. Due to the high level of noises in the original imaging data, accurate estimation and denoising procedures are largely needed. By pooling the estimates of the covariance matrices obtained from different sliding windows together and taking the structure of matrices into account, our newly proposed model can help improve the estimation accuracy of these dynamic covariance matrices. 

\section{Method}\label{method}
\subsection{Model}
Suppose we observe a set of 2D-images and some scalar predictors from $ n $ study subjects. Let  $ Y_i $ be a $p\times q $ matrix representing the 2D-image from the $i$th subject, and $ X_i=(x_{i1},\ldots,x_{is})^T $ be an $ s\times 1 $ vector denoting the scalar covariates of interest (e.g., time and disease predictors). We propose the following nonparametric matrix response model,
\begin{equation}\label{model}
E(Y_i|X_i)=g(X_i),
\end{equation}
where $ g(\cdot): \mathbb{R}^{s} \rightarrow \mathbb{R}^{p \times q}$ is a nonparametric matrix-valued function that quantifies the nonlinear relationship between (each pixel of) $Y_i$ and $X_i$. Since $g(x)$ is a $p \times q$ matrix, we will also impose a structure constraint on $g$ for scientific interpretability and regularization purpose. 

Our goal is to estimate the nonparametric function $ g $. A commonly used estimator is the Nadaraya-Watson estimator for matrix data, which can be written as 
\begin{equation}\label{naive}
\widehat{g}_{\text{nw}}(x)=\frac{\sum_{i=1}^n K_H(x - X_i)Y_i}{\sum_{i=1}^n K_H(x - X_i)},
\end{equation}
where $ K_H(\cdot)=\frac{1}{|H|}K(H^{-1}\cdot) $, $ K(\cdot) $ is a kernel density function, and $ H = \text{diag}(h_1,h_2,\cdots,h_s) $ is a bandwidth matrix.  It is often assumed that $h_1  = \cdots = h_s = h$ for computational convenience.

However, the Nadaraya-Watson estimator is a ``naive'' estimator in our case because it does not utilize the underlying structure of the matrix response $ Y_i $. In particular, the estimator in (\ref{naive}) can also be obtained by vectorizing $Y_i$, applying the Nadaraya-Watson estimator for the vectorized data, and transforming the estimator back to a matrix. To account for the matrix structure, we take another look at the estimator in (\ref{naive}), which can be obtained by solving the following optimization,
\begin{equation}\label{OLS}
\widehat{g}_{\text{nw}}(x)={\rm argmin}_{Y} \sum_{i=1}^n K_H(x-X_i)\|Y_i-Y\|^2_F, 
\end{equation}
where $ \| \cdot \|_F $ is the Frobenius norm of a matrix.  

To further exploit the underlying structure of the 2D response, we introduce a penalty on $Y$ and propose to solve 
\begin{equation}\label{nuclearobjective}
\widehat g(x)={\rm argmin}_{Y}\left\{\frac{1}{2n} \sum_{i=1}^n K_H(x-X_i)\|Y_i-Y\|^2_F+\lambda_n \|Y\| \right\},
\end{equation}
where $ \lambda_n$ is the tuning parameter and $ \|\cdot\|$ is some norm of a matrix. Possible choices are nuclear norms, total variation norms, and their combination; and each of those norms will have different regularization effects on the image outcomes. For this paper, we mainly focus on the nuclear norm regularization for illustration, i.e., writing $\|Y\|$ as $ \|Y\|_*$, which is defined as the sum of all singular values of the matrix $Y$. The nuclear norm is very popular in 2D-image denoising \citep{gu2014weighted,gao2018regularized}. The underlying true 2D-image is often of low rank or approximately low rank, and the nuclear norm regularization can help recover the low rank structure  given a noisy image \citep{Chen2013}. In our case, $ \widehat g(x) $ can be regarded as an image estimate at the point $ x $, and therefore the penalty $ \|Y\|_* $ can push for a low rank representation of the image estimate. 


In Section 6 of Supporting Information, we show that solving (\ref{nuclearobjective}) is equivalent to solving
\begin{equation}\label{nuclearobjective1}
\begin{split}
 \widehat{g}(x)
={\rm argmin}_{Y}\left\{\frac{1}{2}\|\widehat{g}_{\text{nw}}(x)-Y\|^2_F+\frac{n\lambda_n}{\sum_{i=1}^n K_H(x-X_i)}\|Y\|_*\right\}.
\end{split}
\end{equation}

The optimization problem (\ref{nuclearobjective1}) can be solved using the following proposition restated from \citet{cai2010singular}.  
\begin{proposition}
Consider the singular value decomposition of a matrix $Y\in  R^{p\times q}$ with rank $r$,
\begin{align*}
Y = U\Sigma V^\ast,\text{\quad} \Sigma = \text{diag}(\{\sigma_j\}_{1\leq j\leq r}),
\end{align*}
where $U$ and $V$ are $p\times r$ and $q\times r$ matrices respectively with orthonormal columns, and singular values $\{\sigma_j\}_{1\leq j\leq r}$ are positive (we only consider these $r$ non-zero singular values given that the rank of $Y$ is $r$). The soft-thresholding operator $D_\tau$ is defined as
\begin{equation}
D_\tau(Y) = UD_\tau(\Sigma)V^\ast, D_\tau(\Sigma) = \text{diag}[\{(\sigma_j - \tau)_{+}\}_{1\leq j\leq r}], 
\end{equation}
where $(\cdot)_{+}$ is the positive part of $(\cdot)$. Then $D_\tau(Y)$ satisfies
\[
D_\tau(Y) = \arg\min_{X} \left\{\frac{1}{2}\|Y - X\|_F^2 + \tau \|X\|_\ast\right\},
\]
where $\|X\|_\ast$ is defined as the nuclear norm of the matrix $X$.
\label{SVD}
\end{proposition} 
By Proposition \ref{SVD}, our estimator in (\ref{nuclearobjective}) can be obtained using the following algorithm. The proof is given in  Section 7 of Supporting Information.

\begin{algorithm}[H]
\label{algorithm}
\caption{Algorithm to solve the optimization problem (\ref{nuclearobjective}).}
{\bf Input}: $\{(X_i, Y_i), 1\leq i\leq n\}, x, H, \lambda_n$.\\
{\bf Step 1}: Perform singular value decomposition of $\widehat{g}_{\text{nw}}(x)=\frac{\sum_{i=1}^n K_H(x-X_i)Y_i}{\sum_{i=1}^n K_H(x-X_i)}$, and denote it by $U \Sigma V^{\ast}$. The diagonal matrix $\Sigma = \text{diag}(\{\sigma_j\}_{1\leq j\leq r})$, where $ \sigma_1\geq \sigma_2\geq \ldots \geq \sigma_r>0 $ with $ r $ being the rank of $ \Sigma $. \\
{\bf Step 2}: Set $\tau =\frac{n\lambda_n}{\sum_{i=1}^n K_H(x-X_i)}$, and calculate the soft-thresholding operator $D_\tau(\Sigma) = \text{diag}(\{(\sigma_j - \tau)_{+}\}_{1\leq j\leq r})$.\\
{\bf Step 3}: Calculate $\widehat{g}(x) =UD_\tau(\Sigma)V^{\ast}$.\\
{\bf Output}: $\widehat{g}(x)$.
\end{algorithm}  


\subsection{Bayesian information criterion}
The optimization problem (\ref{nuclearobjective}) involves two tuning parameters, the bandwidth $ h $ and the regularization parameter $ \lambda $.  The choices of these two parameters are critical as they control the temporal smoothing level and the spatial low-rank level, respectively. In this paper, we derive a Bayesian information criterion (BIC) to select them. Define $\widetilde\lambda=\frac{n\lambda_n }{\sum_{i=1}^n K_H(x-X_i)}$ and $\widehat Y_i(\widetilde\lambda)= \widehat{g}(X_i)$. Without loss of generality, we assume $ p\geq q $ and denote the singular values of $\widehat Y_i(\widetilde\lambda)$ by $b_{i1}(\widetilde\lambda)\geq \cdots\geq b_{iq}(\widetilde\lambda)\geq 0$. From Algorithm \ref{algorithm}, it can be seen that the singular values of $\widehat Y_i(\widetilde\lambda)$ are corresponding truncated singular values of $\widehat{g}_{\text{nw}}(X_i)$.
Since we are considering a least squared error loss in \eqref{nuclearobjective}, the BIC can be defined as
\begin{equation}
\BIC(\widetilde{\lambda}) = npq\log\left\{\frac{1}{npq}\sum_{i =1}^n \|Y_i - {\widehat Y}_i(\widetilde{\lambda})\|_F^2\right\} + \log(npq)\text{df}(\widetilde{\lambda}),
\label{BIC}
\end{equation}
where $\text{df}(\widetilde{\lambda})$ can be estimated based on the result in the following proposition. The proof of the proposition is deferred to the Supporting Information. 
\begin{proposition}\label{prop:degreefreedom}
Denote $\widehat{g}_{\text{nw}}(X_i)$'s singular values by $\sigma_{i1}\geq \sigma_{i2} \geq\cdots\geq \sigma_{ir_i} > 0$ and $\sigma_{ik} = 0$ for $k > r_i$. An unbiased estimator of the degree of freedom $df(\widetilde{\lambda})$ is 
\begin{align}
& \widehat{df}(\widetilde\lambda)= K_H(0)\sum_{i = 1}^n \frac{\widehat{df}_i(\widetilde{\lambda})}{\sum_{j=1}^n K_H(X_i - X_j)}, \text{\quad where} \nonumber \\
& \widehat{df}_i(\widetilde{\lambda}) = \sum_{k=1}^q { 1}_{\{b_{ik}(\widetilde{\lambda})>0\}}\Big\{ 1 + \sum_{1\leq j\leq p, j\neq k, k \leq r_i} \frac{\sigma_{ik}(\sigma_{ik}-\widetilde{\lambda})}{\sigma_{ik}^2 - \sigma_{ij}^2} + \sum_{1\leq j\leq q, j\neq k, k \leq r_i} \frac{\sigma_{ik}(\sigma_{ik}-\widetilde{\lambda})}{\sigma_{ik}^2 - \sigma_{ij}^2}\Big\}. \label{df_lambda}
\end{align}
\end{proposition}
\section{Theory}\label{theory}
In this section, we present theoretical results of the estimation procedure in \eqref{nuclearobjective}, including a risk bound of the regularized estimator and a rank consistency result. Denote the strength of regularization as $\lambda_n$ and the true response as $g(X)$ given covariates $X$. Assume $g(X)$ has unknown rank $r$ and denote the global minimizer of (\ref{nuclearobjective}) by $\widehat{g}(X)$. For any two sequences of real numbers $a_n$ and $b_n$, we write $a_n \asymp b_n$ if there exist  universal positive constants $C_1$ and $C_2$ such that $C_1 b_n \leq a_n \leq C_2 b_n$. We define $a \vee b = \max(a,b)$ and $a \wedge b = \min(a,b)$ for any $a,b \in \mathbb{R}$. With a little abuse of the notation, we use $C$ to denote a universal constant whose value may change in different context but does not affect the results. For a matrix $A$ and a sequence of real numbers $a_n$, we write $A = O_p(a_n)$ or $A = o_p(a_n)$ if every element of $A$ is $O_p(a_n)$ or $o_p(a_n)$. 

Let $g_{jk}(x)$ be the $(j,k)$-th component of $g(x)$. We make the following assumptions:
\begin{assumption}
We assume that
$|g_{jk}(x) - g_{jk}({ y})| < C\|{ x - y}\|^{\alpha_2}$ with $\alpha_2 > 0$, $1\leq j \leq p, 1\leq k \leq q$ for any $\|{ x -y}\| < \delta$ and some $C > 0$, when $\delta > 0$ is sufficiently small. We also assume that the pixels in the residual matrix, $ E_i = Y_i - g(x_i)$, are independent and identically distributed (i.i.d) for every $i=1,\ldots,n$.  
\label{assump:g_smooth}
\end{assumption}
\begin{assumption}
Assume that $n pq h^{2\alpha_2 +s} \rightarrow \infty$, $nh^{2s} \rightarrow \infty$, and $ pq h^{2\alpha_2} \rightarrow 0$ as $n \rightarrow \infty$.
\label{assump2}
\end{assumption}
\begin{assumption}\label{ap1}
We assume that the kernel function $K(\cdot)$ is bounded on $\mathbb R^s$. In other words, there exists a universal constant  $k_\text{max} > 0$ (i.e, $k_{\max}$ does not depend on $x$) such that $K(x) \leq k_{\max}$ and  $K_H(x) \leq h^{-s} k_\text{max}$ for every $x \in \mathbb{R}^s.$ Moreover, we assume there exist universal constants $C_f, c_f > 0$ such that the density function of covariate $x$ satisfies $c_f \leq f(x) \leq C_f$ for every $x$.
\end{assumption}

\begin{assumption}
Assume that $n^{-1/2}(p \wedge q) \to 0$, $(pq)^{1/2}h^{\alpha_2}(p \wedge q)^{1/2}\lambda_n^{-1} \to 0$, $\lambda_n (p \wedge q)^2 \to 0$, and $ \frac{ pq (p \wedge q)}{n\lambda_n^2}  \to 0$ as $n \rightarrow \infty$.
\label{a2}
\end{assumption}

Assumption \ref{assump:g_smooth} assumes the $\alpha_2$-smoothness for each element of the nonparametric function $g(x)$. This assumption is commonly used in multivariate function estimation literature such as \citet{scott2015multivariate}. It is possible to extend the results for anisotropic case in the future work based on the techniques developed in this paper. The i.i.d assumption for the pixels in the residual is needed for showing consistency result, and we have conducted a sensitivity analysis in the simulation to show that our proposed method is robust to the violation of this assumption.  Assumption \ref{assump2} is required for estimation consistency as it imposes constraints on the matrix size and the bandwidth. Assumption \ref{ap1} is satisfied for most kernel density functions. The boundedness condition for the density function of $x$ will hold if $x$ is defined on a compact support and the density is continuous. Assumption \ref{a2} is needed for rank consistency. With these assumptions, we can state the two main theorems of this paper. Their proofs are given in the Supporting Information. 

\begin{theorem}
\label{thm:consistency}
Suppose that Assumptions \ref{assump:g_smooth}--\ref{ap1} hold. We consider two cases for $p$ and $q$ (e.g., whether they diverge or not). 

(1) If both $p$ and $q$ are fixed, let 
$h \asymp (\frac{\log n}{n})^{\frac{1}{2\alpha_2+s} \wedge  \frac{1}{2s}}$ and $\lambda_n \asymp  h^{\alpha_2}$, then with probability tending to 1, 
\begin{align*}
\|\widehat{g}(x) - g(x) \|_F^2 \leq C  r \left(\frac{\log n}{n}\right)^{ \frac{2\alpha_2}{2\alpha_2+s} \wedge \frac{\alpha_2}{s}}.
\end{align*}

(2) If $p \vee q \rightarrow \infty$ and $p \vee q = o\left( n^{\frac{\alpha_2}{2\alpha_2 +s}} \wedge (\frac{n}{\log n})^{\frac{\alpha_2}{2s}}\right)$, then by letting $h \asymp  n^{-\frac{1}{2\alpha_2 + s}} \vee (\frac{\log n}{n})^{\frac{1}{2s}}$ and $\lambda_n = h^{\alpha_2} (p q)^{1/2}$, with with probability tending to 1, we have
\begin{align*}
\|\widehat{g}(x) - g(x) \|_F^2 \leq C pqr \left\{n^{-\frac{2\alpha_2}{2\alpha_2 + s}} \vee (\frac{\log n}{n})^{\frac{\alpha_2}{s}}\right\}.
\end{align*}
\end{theorem}
Note that the risk bound involves two quantities $n^{-\frac{2\alpha_2}{2\alpha_2 + s}}$ and  $(\frac{\log n}{n})^{-\frac{\alpha_2}{s}}$. As the number of predictors $s$ increases, it becomes more difficult to estimate $g(x)$. Meanwhile, $h^s$ is involved  when proving strongly restricted convexity of the loss function. A larger value of $s$ indicates smaller probability of the loss function being strong restricted convex.
In contrast, $\alpha_2$ describes the smoothness of $g(x)$. A larger $\alpha_2$ leads to a smaller risk bound and a faster convergence rate. The rate here holds uniformly for any $x$ satisfying Assumption 3; and the true rank of matrix $r$ does not have to be finite for case (2).  
 
If $p$ and $q$ are fixed and $s \le 2\alpha_2$, the optimal bandwidth $h$ can be chosen arbitrarily close to $n^{-\frac{1}{2\alpha_2 + s}}$, which leads to the same convergence rate (with additional logarithmic factor) for estimating an $\alpha_2$-smooth, $s$-dimensional function without regularization. When $\max (p, q) \rightarrow \infty$, if we further assume $s \leq \alpha_2$ and choose $nh^{2\alpha_2 + s} \asymp \frac{(\sqrt p + \sqrt q)^2}{pq}$ and $\lambda_n \asymp (pq)^{1/2} \left\{\frac{(\sqrt p + \sqrt q)^2}{npq}\right\}^{\frac{\alpha_2}{2\alpha_2 + s}}$,
as we let $\lambda_n \rightarrow 0$ and $nh^{2s} \rightarrow 0 $, we obtain $\max(p,q) = o(n^{\frac{\alpha_2}{\alpha_2 + s}})$. This is the necessary condition for $\widehat{g}(x)$ being consistent.  The assumption $s \leq \alpha_2$ rules out the case where there are too many covariates in the model.

Next we present the rank consistency result. We consider three general cases for different values of $p$ and $q$ (e.g., whether they diverge or not) and discuss the corresponding choices of $\lambda_n$ and $h$ as follows, 
\begin{itemize}
\item[(C1)] If both $p, q$ are fixed, we can choose $h \asymp (\frac{\log n}{n})^{\frac{1}{2\alpha_2+s} \wedge \frac{1}{2s}}$ and   $\lambda_n \asymp  h^{\alpha_2}\log n$. 
\item[(C2)] If $p \wedge q $ is  finite, and $p \vee q \to \infty $ satisfying $(\log n)^2 (p \vee q) = o\left\{  (\frac{n}{\log n})^{\frac{\alpha_2}{s}} \wedge n^{\frac{2\alpha_2}{2\alpha_2+s}}\right\}$, then we choose $h =  (\frac{\log n}{n})^{\frac{1}{2s}} \vee n^{-\frac{1}{2\alpha_2 + s}}$ and  $\lambda_n \asymp (p \vee q)^{1/2} h^{\alpha_2}\log n$.  
\item[(C3)] If $p\asymp q$, and $p \to \infty$, then we let $h \asymp  (\frac{\log n}{n})^{\frac{1}{2s}} \vee n^{-\frac{1}{2\alpha_2 + s}}$ and $\lambda_n \asymp p^{\frac{3}{2}} h^{\alpha_2} (\log n)$. In addition, we assume $(\log n)^{\frac{2}{7}} p = o(h^{-\frac{2\alpha_2}{7}}) $.
\end{itemize}

\begin{theorem}\label{thm:rank consistency}
Suppose that Assumptions \ref{assump:g_smooth}--\ref{a2} hold, and one of the cases in (C1)--(C3) holds, then $\widehat{g}(x)$ is consistent and rank consistent, i.e.,
\begin{equation*}
P\left[\text{rank}\{\widehat{g}(x)\} = \text{rank}\{g(x)\}\right]\to 1,\text{\quad as } n \to \infty.
\end{equation*}
\end{theorem}
It can be seen that rank consistency requires stronger assumptions on $p, q$ compared with those in Theorem \ref{thm:consistency}. For instance, the desired $\lambda_n$ is much larger than the one from the previous theorem. Meanwhile, $pq$ is not allowed to be greater than $n$ for rank consistency. Assumption \ref{a2} is an extra condition needed for rank consistency but not for estimation consistency. The rank consistency result holds uniformly for any $x$ satisfying Assumption \ref{ap1}. 

\section{Simulation}\label{simulation}
In this section, we evaluate the empirical performance of our method and other competing methods. We consider both univariate and multivariate $ X $, different nonparametric functions and different correlation structures of the random error  $ E_i$, where $E_i = Y_i - g(x_i)$.  
\subsection{Univariate predictor}
\label{univariate}
{\bf Setting I}: We set the dimensions of the image $ p=q=64 $, and set the $(j,k)$-th element of the nonparametric function $g(x)_{jk} = \{\sin(10\pi x) + \cos(10\pi x)+0.1(j+k)\}*B_{jk}$, $1\leq j, k\leq 64$, where $0\leq x \leq 1 $ and $ B_{jk} $ is the $(j,k)$-th element of the true signal $B$. The true signal $B$ is generated from a 64-by-64 image, where we consider three shapes: a cross, a square and a T-shape. We have plotted the true shapes in Figure 1(a)(b)(c), where we assign $B$ a value of 5 for black regions and 0 for white regions. The sample size is set at $n = 200, 500$. The covariates $\{x_i\}, i=1, 2,\dots,n$ are equally spaced on [0,1]. The response $Y_i $ is generated from $Y_i=g(x_i)+E_i,$ where $\text{vec}(E_i)$'s are i.i.d $N(0,I_{pq})$. The optimal bandwidth $h$ and $\lambda$ are selected by BIC. For the kernel function, we use the standard gaussian kernel density defined as $K(x) = \exp(-x^2/2)/\sqrt{2\pi}$. We compare our method with the naive Nadaraya-Watson estimator and the Lasso estimator, where the Lasso estimator is obtained by solving the following optimization problem
\begin{equation}
\widehat g_{\text{lasso}}(x)={\rm argmin}_{Y}\left\{\frac{1}{2n} \sum_{i=1}^n K_H(x-X_i)\|Y_i-Y\|^2_F+\lambda_n \|Y\|_1\right\},
\label{Lasso}
\end{equation}
where $ \|Y\|_1 $ is defined as the sum of the absolute values of all the elements of the matrix $ Y $. We also use the BIC as defined in (\ref{BIC}) to choose the tuning parameter for Lasso. Here the degree of freedom can be obtained by the chain rule as
\begin{align*}
\widehat{\text{df}} &= \sum_{i=1}^n\text{tr}\left(\frac{\partial \text{vec}(\widehat Y_i)}{\partial \text{vec}(Y_i)}\right)= \sum_{i=1}^n\sum_{j=1}^{p}\sum_{k=1}^{q}\frac{\partial \widehat Y_{ijk}}{\partial \widehat g_{ijk(nw)}}\frac{\partial \widehat g_{ijk(nw)}}{\partial Y_{ijk}} \\
& =
\sum_{i=1}^n\frac{K_H(0)}{\sum_{j=1}^n K_H(X_i - X_j)}\|\text{sign}(\text{vec}(\widehat Y_i))\|_1,
\end{align*}
where we have used the fact that $\frac{\partial \widehat Y_{ijk}}{\partial \widehat g_{ijk(nw)}} = |\text{sign}(\widehat Y_{ijk})|$. 

In each replication, we generate $n$ samples as the training set and another 500 samples as the test set. We report the integrated error $\int_x \|\widehat Y(x)-Y(x)\|_F^2dx$, which can be approximated by $\frac{1}{500}\sum_{i=1}^{500} \|\widehat Y(x_i^{test})-Y(x_i^{test})\|_F^2$. Table \ref{tb:q_AR}  shows the average integrated test error by our method, naive Nadaraya-Watson estimator and Lasso estimator based on 100 Monte Carlo replicates. We also report the average selected rank by our method using BIC, defined as $\frac{1}{n}\sum_{i=1}^n\text{rank}\{\widehat Y(x_i)\}$. 
 
\begin{figure}
\centering
\subfigure[]{\includegraphics[width=0.25\linewidth]{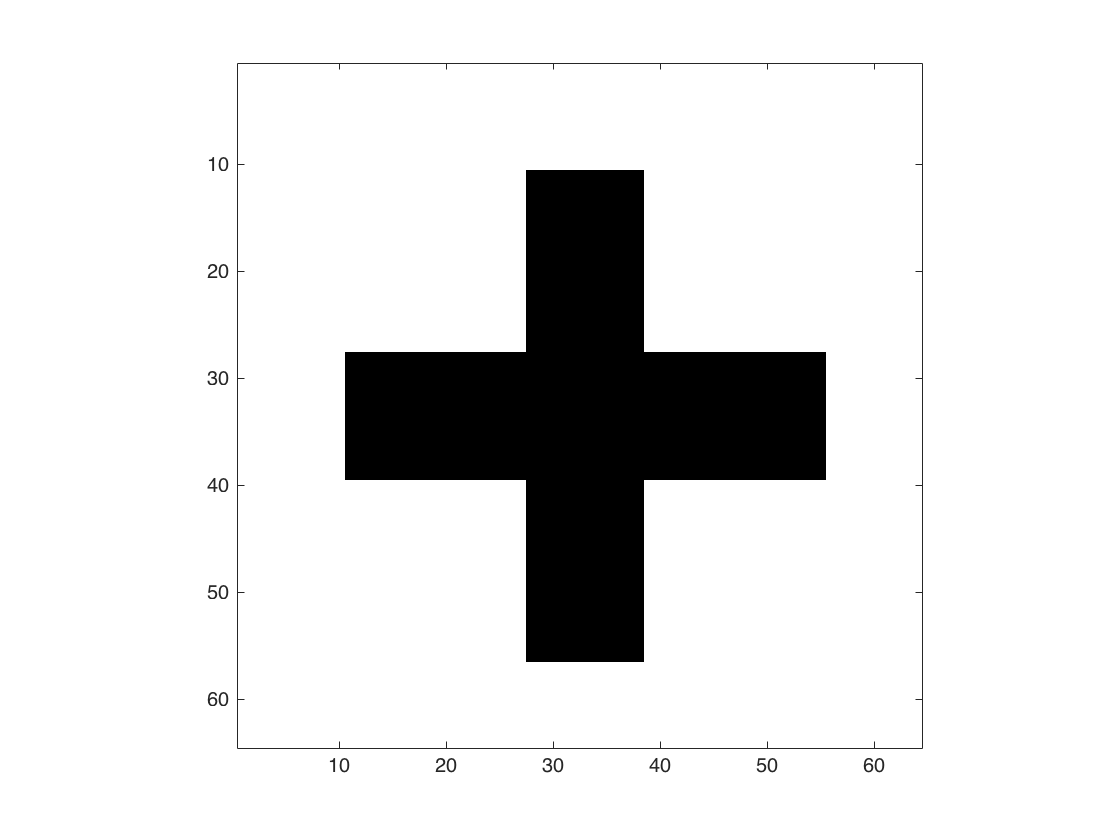}}
\subfigure[]{\includegraphics[width=0.25\linewidth]{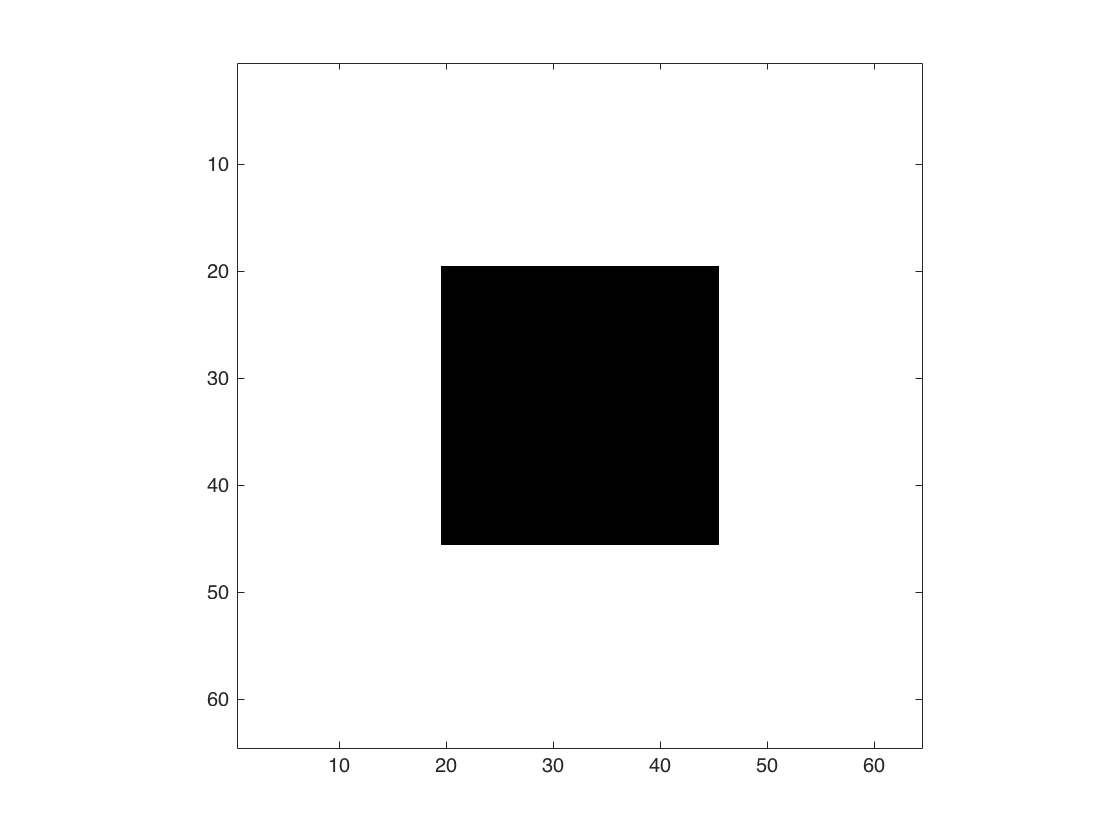}}
\subfigure[]{\includegraphics[width=0.25\linewidth]{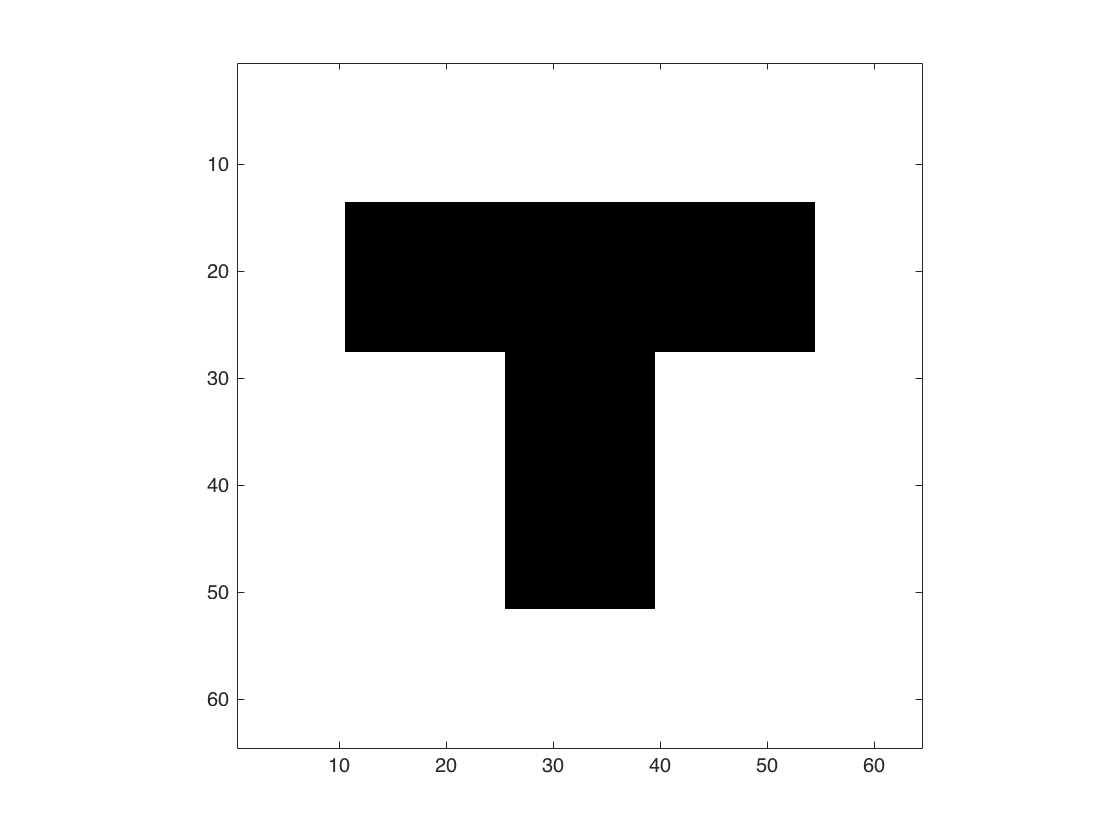}} \\
\subfigure[]{\includegraphics[width=0.25\linewidth]{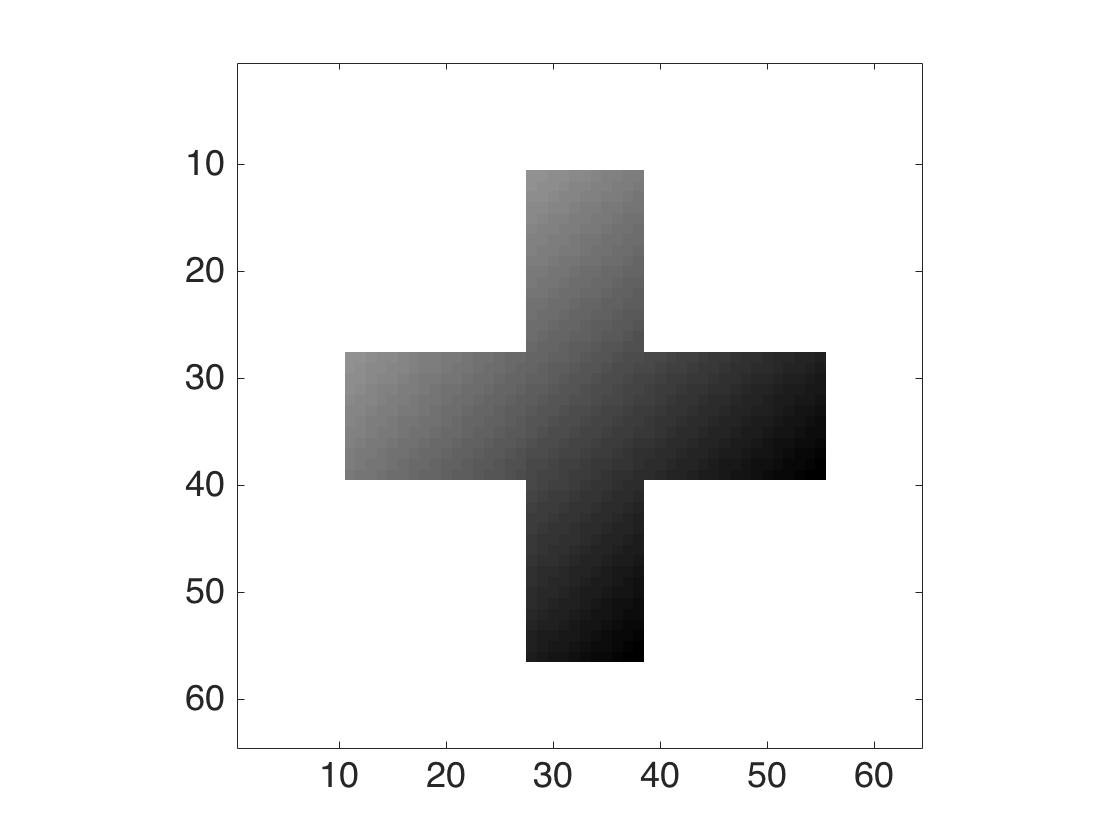}}
\subfigure[]{\includegraphics[width=0.25\linewidth]{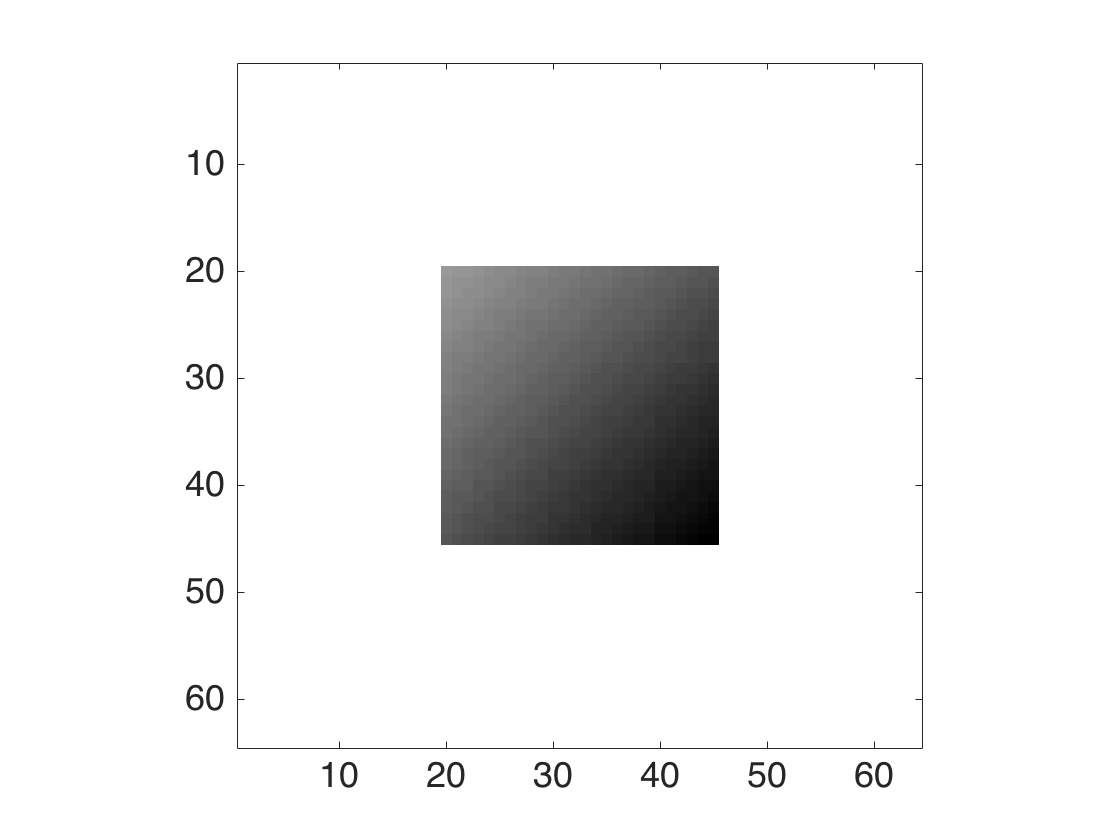}}
\subfigure[]{\includegraphics[width=0.25\linewidth]{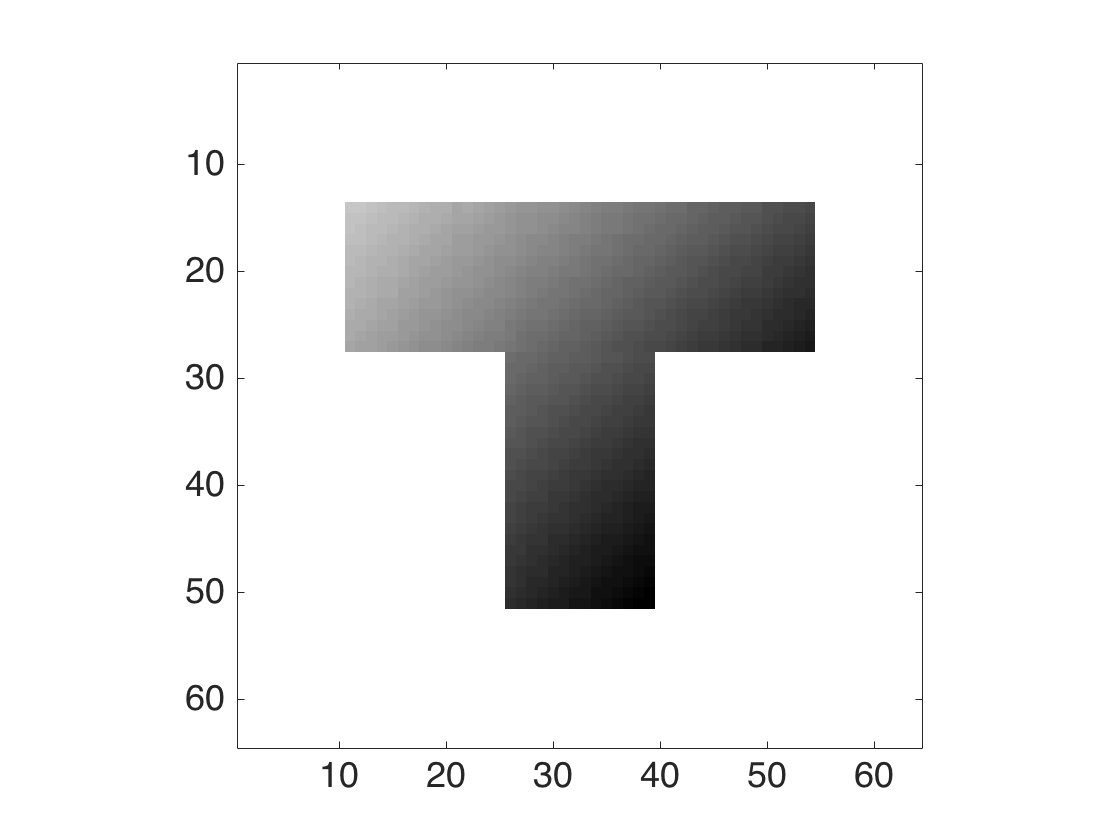}}
\caption{(a)-(c): true signals, (d)-(f): recovered signals}
\label{fig:three-shape}
\end{figure}

From the results, we can see that our method performs better than Nadaraya-Watson estimator and Lasso estimator in all cases. In addition, our method can estimate the true rank of the image accurately. We have plotted the recovered signals from one randomly selected Monte Carlo study in Figure \ref{fig:three-shape}(d)(e)(f), and our method manages to recover the true signals very well.

\begin{table}
\def~{\hphantom{0}}
\caption{Simulation results: mean of integrated test error and associated standard errors obtained from our
method, NW estimator, and Lasso, the average selected rank and true rank are reported for three different shapes $ B $. The results are based on 100 Monte Carlo replications. }{%
\begin{tabular}{lllccccc}  \hline
Setting I & $n$ & Shape&Our method &NW &Lasso& Selected rank &True rank\\[5pt]
& &Cross&4458 (0.70)&6024 (0.96)&5148 (0.81)&3.53 (0.004)&4\\
& 200&Square&4288 (0.82)&6107 (0.99)&4875 (0.83)&2.00 (0.000)&2\\
& &Tshape&4472 (0.76)&6094 (0.98)&5193 (0.92)&3.22 (0.006)&4\\
&&&&&&&\\
&& Cross&4306 (0.41)&5009 (0.52)&4560 (0.48)&3.99 (0.000)&4\\
& 500 &Square&4186 (0.41)&4803 (0.48)&4440 (0.45)&2.01 (0.000)&2\\
& &Tshape&4255 (0.60)&5042 (0.51)&4579 (0.49)&3.52 (0.007)&4\\ \hline
Setting II & $n$ & Shape&Our method &NW &Lasso& Selected rank &True rank\\[5pt]
 & &Cross&4656 (2.23)&6120 (3.47)&5528 (3.49)&6.64 (0.009)&4\\
&   200&Square&4463 (2.31)&5785 (3.27)&5169 (4.72)&4.42 (0.009)&2\\
& &Tshape&4667 (2.49)&6017 (3.65)&5591 (3.74)&6.52 (0.008)&4\\
& &&&&&&\\
& & Cross&4403 (1.23)&5240 (1.96)&4769 (1.71)&6.90 (0.008)&4\\
&   500 &Square&4296 (1.10)&5036 (1.64)&4692 (1.68)&4.76 (0.000)&2\\
&  &Tshape &4404 (1.21)&5057 (1.60)&4797 (1.69)&6.75 (0.008)&4\\ \hline
 Setting III & n &Shape&Our method &NW &Lasso& Selected rank &True rank\\[5pt]
&  &Cross&4711 (0.86)&7021(1.17)&5759(1.01)&4.03 (0.002)&4\\
 &  200&Square&4518 (0.80)&6723(1.09)&5326(1.05)&2.04 (0.002)&2\\
 & &Tshape&4719 (0.83)&7137 (1.11)&5829(1.15)&4.34 (0.005)&4\\
 &&&&&&&\\
 & &Cross&4647 (0.49)&5562 (0.62)& 4843 (0.48) &4.84 (0.006)&4\\
& 500 &Square&4281 (0.42)&5506 (0.58)&4649(0.45)&2.01 (0.001)&2\\
&& Tshape&4376 (0.45)&5620 (0.59)&4875 (0.49)&4.12 (0.002)&4\\ \hline
Setting IV & $n$ &Shape&Our method &NW &Lasso& Selected rank &True rank\\[5pt]
 & &Cross&4894 (2.58)&7164 (3.89)&5804 (4.13)&5.40 (0.008)&4\\
& 200&Square&4642 (2.43)&6858 (3.57)&5360 (4.0)&3.14 (0.009)&2\\
&  &Tshape&4910 (2.66)&7283 (3.91)&5880 (4.36)&5.36 (0.008)&4\\
 &&&&&&&\\
 & &Cross&4779 (1.73)&5687 (2.38)& 5067 (1.86) &6.38 (0.008)&4\\
& 500 &Square&4574 (1.51)&5614 (2.22)&4815 (1.62)&4.12 (0.001)&2\\
&& Tshape&4797 (1.55)&5745 (2.09)&5080 (4.72)&6.34 (0.008)&4\\ \hline
 \end{tabular}}
 \label{tb:q_AR}
\end{table}

{\bf Setting II}: In this setting, we consider the case where the errors $\text{vec}(E_i)$ are correlated across different subjects $i$'s and the pixels within the same random error matrix $ E_i $ are also correlated. Define $ {\bf e}=(\text{vec}(E_i)^{\T}, \ldots, \text{vec}(E_n)^{\T})^{\T}\in \mathbb{R}^{pqn}$. We assume $ {\bf e}\sim N(0, \Sigma) $, where $\Sigma= \Sigma_1\otimes \Sigma_2\in \mathbb{R}^{pqn\times pqn} $. Here $ \Sigma_1$ is a $ n\times n $ matrix representing the correlation between different subjects $ 1\leq i\leq n$, $ \Sigma_2 $ is a $ pq\times pq $ matrix representing the correlation among different pixels of the 2D image, and $\otimes$ is the Kronecker product. This decomposition of $\Sigma$ is often referred to as the separability of the covariance matrix, which was studied in the literature such as \citet{DeMunck2002,Dawid1981}. For $\Sigma_1$, we assume it has  a subject-wise 1D autoregressive structure. In particular, we set the $(i_1,i_2)$-th element of $\Sigma_1 $ as $ 0.5^{|i_1-i_2|}$ for $1\leq i_1,i_2\leq n $. For $ \Sigma_2$, we assume it is incorporated with a pixel-wise 2D autoregressive structure. Specifically, we set the $ (j_1+(k_1-1)q, j_2+(k_2-1)q)$-th element of $\Sigma_2$ as $0.5^{|j_1-j_2|+|k_1-k_2|}$ for $1\leq j_1,j_2\leq p$ and  $1\leq k_1,k_2\leq q$. The average integrated test errors by three methods and the average selected rank of our method are summarized in Table \ref{tb:q_AR}. From the results, we can see that our methods still outperforms Nadaraya-Watson estimator and Lasso estimator in all cases. Compared with independent error case, we may over select the rank a bit, possibly due to the error correlations, however, the average integrated errors are still at the similar levels for both cases.

\subsection{Multivariate predictors}
{\bf Setting III}: We consider shapes of the image with the same pixel value as setting I. We set the $(j,k)$-th element of the nonparametric function $g(x)_{jk} = \{\sin(2\pi\|x\|)+\cos(2\pi\|x\|)+0.5(j+k)\}*B_{jk}$, $x\in [0,1]\times[0,1]$, $1\leq j,k \leq 64$, where we consider the same three shapes of the true image $ B $ and $\|x\|$ is the $l_2$-norm of $x$. The random error $\text{vec}(E_i)$'s are i.i.d. $N(0,I_{pq})$. The covariates $x_i$ consist of a set of $\{x_{jk}\}$, $ 1\leq j\leq 20,1\leq k\leq 25$, that are equally spaced on $[0,1]\times[0,1]$. The sample sizes $n = 200, 500$ are considered, and the multivariate Gaussian kernel defined as $K(x) = \exp(-\|x\|^2/2)/2\pi$ is used.  In each Monte Carlo simulation, we generate $n$ samples as the training set and another 500 samples as the test set. We report the average integrated test error obtained by our method, the NW estimator and Lasso estimator and the average selected rank of our method based on 100 Monte Carlo replicates in Table \ref{tb:q_AR}. 

{\bf Setting IV}: We consider same setting as Setting III except that the random error $\text{vec}(E_i)$'s are correlated across different $i$'s and the pixels within the same random error matrix $ E_i $ are also correlated. The random error $\text{vec}(E_i)$'s are generated the same as Setting II. The average integrated test errors by three methods and the average selected rank of our method are summarized in Table \ref{tb:q_AR}. 
The findings in the multivariate case (Settings III and IV) are consistent with the ones in the univariate case. The simulation results in this section confirm the excellent performance of the proposed nonparametric estimation procedure. 

We also conduct a sensitivity analysis to evaluate the performance of our method when the underlying true signal matrix does not satisfy the low rankness assumption. For reach of the four simulation settings, we add $.1$ to every diagonal element in $\Sigma$, the matrix of singular values obtained from the SVD for the true 2D image signal matrix $\text{E}(Y)$. The true signal matrix then has a full rank of $64$. We evaluate the performance of our proposed method and competitive methods, and summarize the estimation error in Section 8 of the Supporting information. We find that our method still has a significantly better performance than that of the NW estimator under all settings, which confirms the benefit of incorporating the matrix structure in the true signal (i.e., the true signal matrix still enjoys a `sparse' structure in the sense that most singular values are not too far away from 0). Comparing to Lasso, our method has a lower estimation error under Settings I and II, but a higher error when the model  becomes more complex (Settings III and IV). Given that the modeling assumption is severely violated, it is fair to conclude that our proposed method has a robust performance under the violation of the low rankness assumption.

We also implement a modified version of our method called by ESS-adjusted method, where we calculate the estimated effective sample size (ESS) based on the true distribution of residuals to replace the original definition of sample size ($npq$) in BIC calculation under simulation setting II and IV, where the generated residuals have a dependence structure. In Table 1 of the Supporting information, we find that the use of ESS indeed help improve the estimation accuracy while the use of original sample size still yields satisfactory results. 

The implementation of ESS adjustment in the simulation is convenient only because we know the true distribution of the residuals. In practice, the residual distribution is usually unknown, and the ESS calculation will then require us to first fit a regression model and then estimate the covariance matrix of the matrix-valued residuals, which itself is a non-trivial problem \citep{Zhang2020}, in order to obtain an estimated ESS. With the new BIC and the selected tuning parameter, the regression estimate will likely to change as well, resulting in a new set of residuals. Therefore, this process may have to be repeated several times to achieve convergence. Developing a computationally efficient ESS adjustment will be an interesting future research direction to pursue. 

\section{Real data application}\label{data}
\subsection{Application to calcium imaging data study}
Calcium imaging has been used as a promising technique to monitor the dynamic activity of neuronal populations. We apply the proposed method to one-photon calcium imaging dataset collected by Ilana Witten's lab at the Princeton Neuroscience Institute \citep{petersen2018scalpel}. Calcium imaging is an important fluorescent microscopy technique regulating a great variety of neuronal processes simultaneously \citep{berridge1998neuronal, andilla2014sparse}. The calcium imaging data can be viewed as a video clip (i.e., a collection of 2D-images recorded at the same frame over a period of time) that presents the location and time of neuron firing \citep{apthorpe2016automatic,petersen2018scalpel}.
Each pixel in a frame is continuous-valued and larger values indicate higher
fluorescent intensities caused by greater calcium concentrations. 
The calcium imaging video we used consists of  3000 frames of size $ 205 \times 226$ pixels sampled at 10 Hz.  An example frame randomly selected from the video is shown in Figure \ref{fig: neuron_curve}(c).

\begin{figure}[h]
\centering
\subfigure[]{\includegraphics[height=4.8cm,trim=0cm 0cm 0cm 0cm,clip]{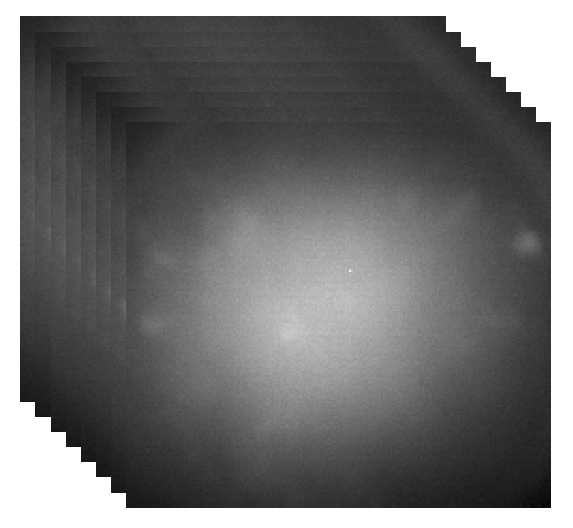}} 
\subfigure[]{\includegraphics[height=5cm,trim=0cm 0cm 0cm 0cm,clip]{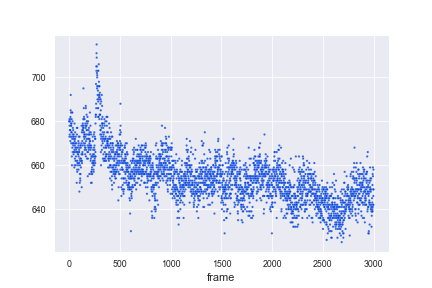}}
\caption{(a) A sequence of frames (b) scatterplot for a fixed voxel of coordinate (200,60) over frames. This figure appears in color in the electronic version of this article, and any mention of color refers to that version. }
\label{fig: bunchframe_scatter}
\end{figure}

\begin{figure}[h]
\centering
\subfigure[]{\includegraphics[height=5cm,trim=1cm 0cm 1cm 0cm,clip]{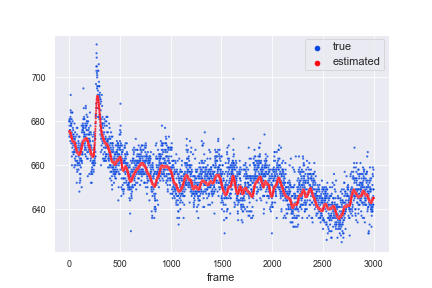}}
\subfigure[]{\includegraphics[height=5cm,trim=1cm 0cm 1cm 0cm,clip]{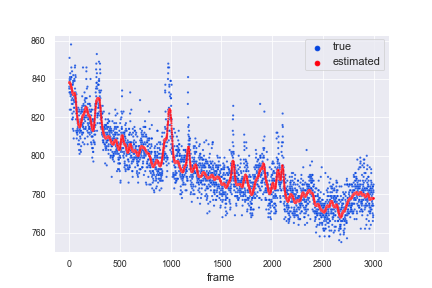}} \\
\subfigure[]{\includegraphics[height=4.2cm,trim=5.3cm 0cm 3cm 0cm,clip]{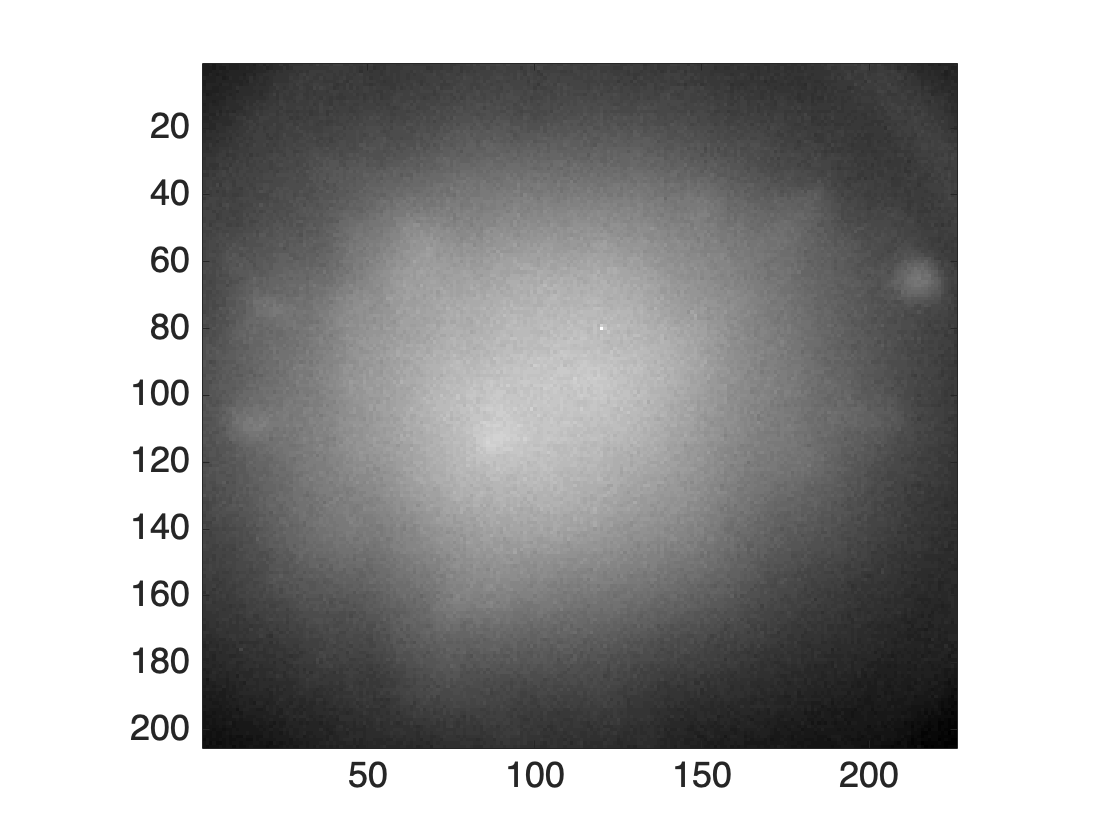}}
\subfigure[]{\includegraphics[height=4.2cm,trim=3cm 0cm 1cm 0cm,clip]{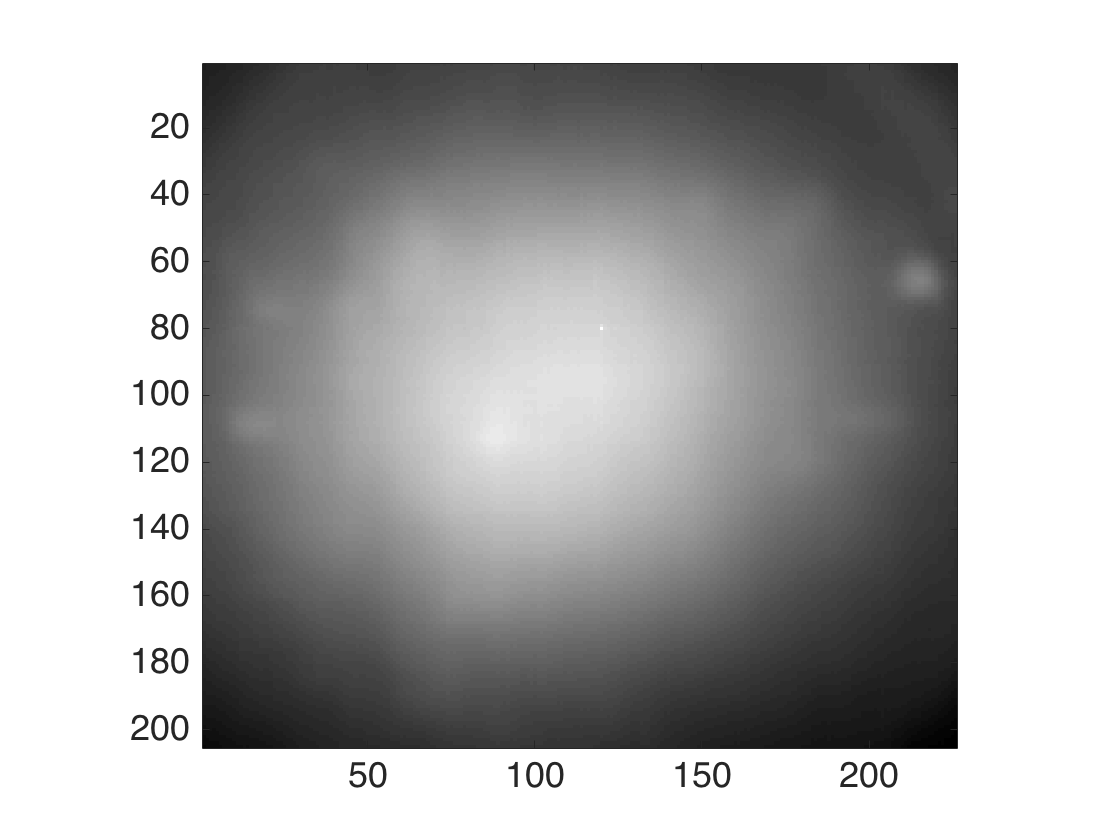}}
\caption{(a) Fitted value for a fixed voxel of coordinate (200,60) over frames  (b) Fitted value for a fixed voxel  of coordinate (60,180) over frames (c) Original 1500th frame (d)  Estimated 1500th frame by our method. This figure appears in color in the electronic version of this article, and any mention of color refers to that version.}
\label{fig: neuron_curve}
\end{figure}

Figure \ref{fig: neuron_curve}(a)(b) give  estimated fluorescent intensities versus true values of two randomly selected pixels across 3000 frames. The optimal bandwidth and regularization parameter are selected by the proposed BIC criteria. The average rank selected by our method is 22.34 (SE=1.34)  across all frames. We also plot estimated images from a randomly selected frame by our method in Figure \ref{fig: neuron_curve}(d).  From that figure, we can see that our method amplifies potential neuron signals, but also weakens those unclear and smaller neurons.

We further evaluate the prediction performance of our method by cross-validation. We compare our method with two nonparametric regression methods: Nadaraya-Watson regression and Lasso defined in \eqref{OLS} and \eqref{Lasso}, respectively. We also compare our method with the low-rank matrix response linear regression (L2RM) method \citep{kong2018l2rm}. The leave-one-out cross-validation error for our method is 2.66 (SE = .63), 2.91 (SE = 1.04) for NW estimator, 2.91 (SE = 1.04) for Lasso estimator, and 5.45 (SE = 3.63) for L2RM. 
It is clear that our method reaches the smallest leave-one-out cross-validation error among four methods. For Lasso estimator, the selected tuning parameter is always zero, hence making the Lasso estimator equivalent to the Nadaraya-Watson estimator. This confirms that the sparsity assumption does not seem plausible for the calcium imaging application, while low rankness is a more reasonable assumption. The linear model L2RM has the highest prediction error, which validates the use of a nonparametric model for the data set. 

\subsection{Application to EEG data study}
We also apply our method to an EEG dataset that was collected from 122 subjects by the Neurodynamics Laboratory to examine the EEG correlates of genetic predisposition to alcoholism. More details about the study can be found in \citet{zhang1995event}. Among the 122 subjects, 77 were alcoholic individuals and 45 were controls. The dataset included voltage values from 64 electrodes placed on each subject's scalps sampled at 256 Hz (3.9- msec epoch) for 1 second. Each subject was exposed to three stimuli: a single stimulus, two matched stimuli, two unmatched stimuli.  For each subject, we use the average of all trials for each subject under single-stimulus condition, which results in a $256\times64$ matrix. Among those 122 subjects, we randomly select one alcoholic individual and one control, and analyze the dynamic functional connectivity among different electrodes across time. The simplest analytical strategy to investigate dynamic functional connectivity consists in segmenting the time courses from spatial locations into a set of temporal windows, inside which their pairwise connectivity is probed. By gathering functional connectivity descriptive measures over subsequent windows, fluctuations in connectivity can be captured. The basic sliding window framework has been applied by the neuroimaging community to understand how brain dynamics that are related to our cognitive abilities \citep{Kucyi2014, Madhyastha2014}, is affected by brain disorders \citep{Sakouglu2010, Jones2012}, or compares to other functional or structural brain measures \citep{Leonardi2013, Tagliazucchi2012, Liegeois2016}. More specifically, we use a moving window of size 100 to calculate a series of covariance matrices along dimension of 256, resulting 157 covariance matrices of size $64\times64$ for each individual. 

\begin{figure}[H]
\centering
\subfigure[]{\includegraphics[height=3.5cm,trim=6cm 0cm 4cm 0cm,clip]{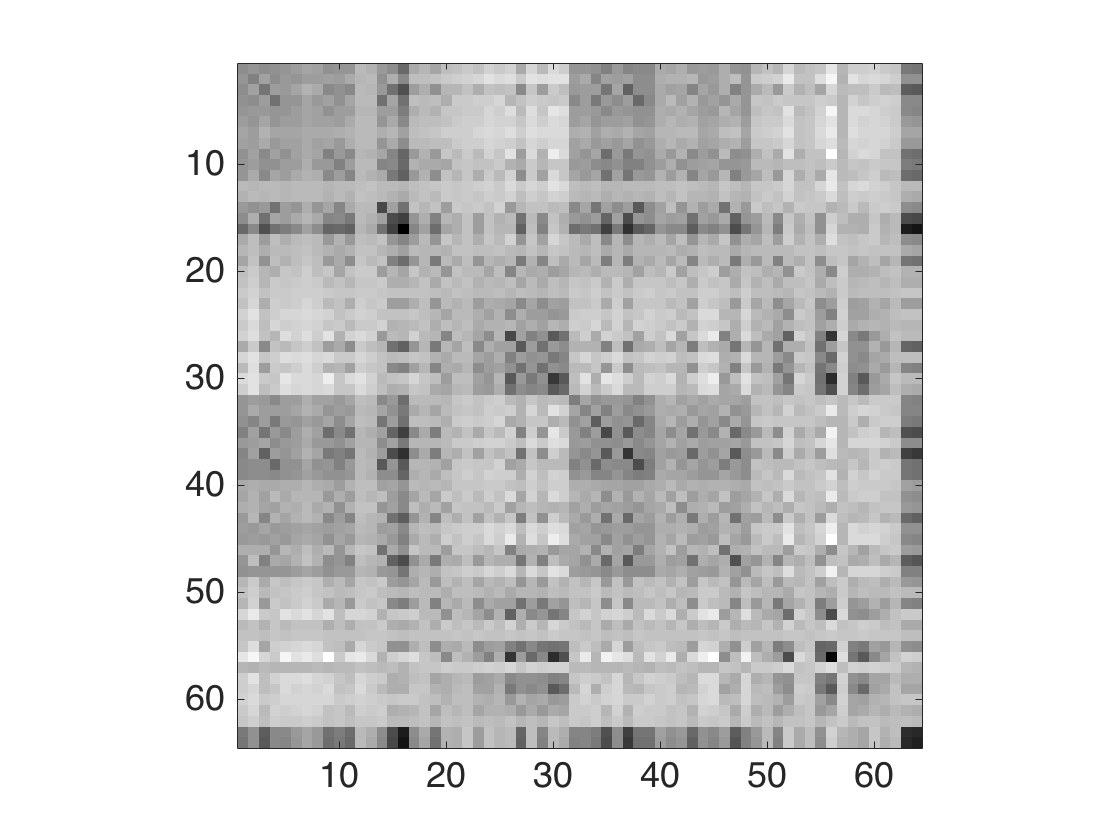}}
\subfigure[]{\includegraphics[height=3.5cm,trim=4cm 0cm 1cm 0cm,clip]{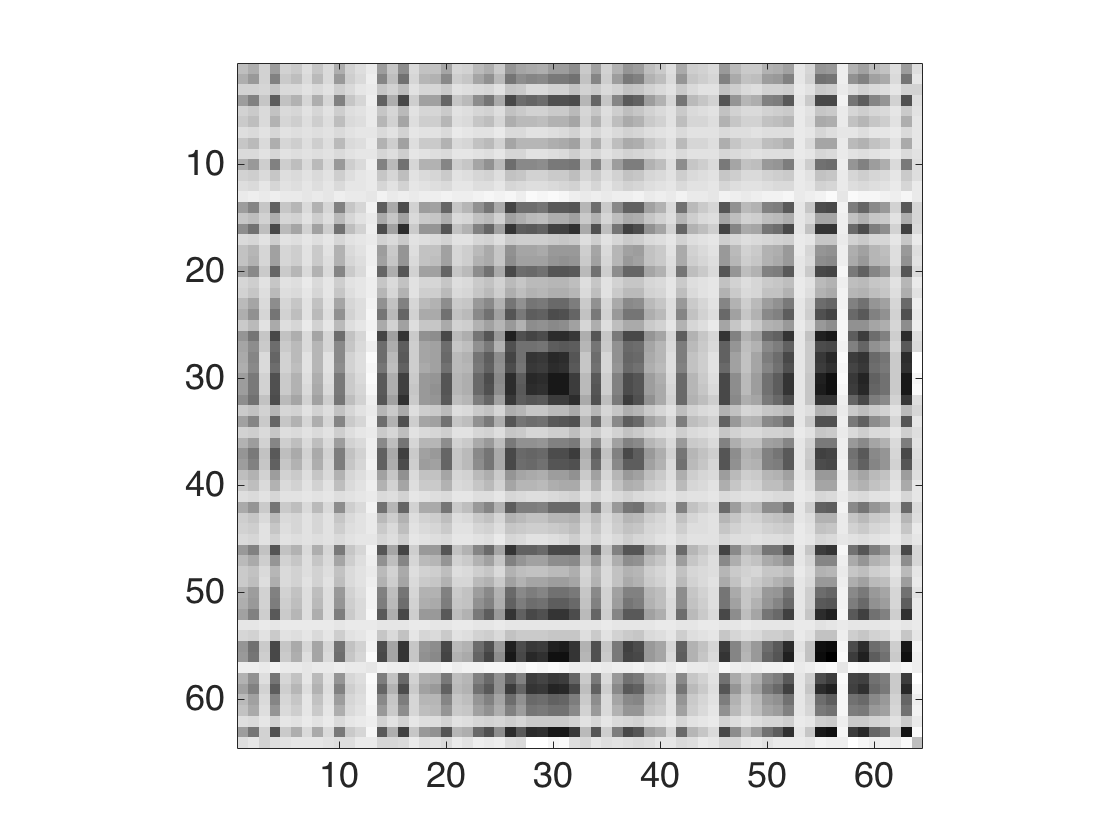}}
\caption{(a) Estimated 10th frame for alcoholic (b) Estimated 10th frame for control}
\label{fig:eeg_matrix}
\end{figure}

We apply the proposed method to analyze the dynamic change of covariance structures over the time in both alcoholic individual and control. In other words, the matrix responses of interest are the dynamic covariance matrices representing the dynamic functional connectivity. The optimal bandwidth and regularization parameter are selected by BIC. Figure \ref{fig:eeg_matrix} shows estimated images of 10th frame by our method for alcoholic individual and control respectively. 
 We observe a significant structural difference in their covariance matrices. Specifically, the alcoholic individual has a more complex covariance structure than that from the control. Moreover, the average selected rank of alcoholic individual is 22.44 (SE = 0.93) compared to 6.82 (SE = 0.87) of control. This can be explained by drastic fluctuation across time in EEG signals of alcoholic individuals compared to stable variation in control. We further evaluate our method by  leave-one-out cross-validation error and compare it with Nadaraya-Watson regression, Lasso and L2RM. As shown in Table \ref{tb:eeg_block_cv}, our method achieves the smallest leave-one-out cross-validation error among three methods. For Lasso estimator, the selected tuning parameter is always zero. In other words, the Lasso estimator is the same as the Nadaraya-Watson estimator for this data application, which implies that the low rankness assumption is a more reasonable assumption than sparsity. We also notice that linear L2RM has a much higher estimation error than the nonparametric methods. This indicates a strong nonlinear pattern in EEG signals for both alcoholic and control subjects. 

\begin{table}[h]
\centering
\def~{\hphantom{0}}
\caption{Leave-one-out cross-validation errors (SE) by three methods for EEG data}{%
\begin{tabular}{lcccc} 
 &Err - Our method&Err - NW&Err - Lasso& Err - L2RM \\
Alcoholic&1.05 (.008)&2.20 (.018)&2.20 (.018)& 908.29 (3.97) \\
Control &9.57 (.391)&21.87 (.75)&21.87 (.75)& 22513.17 (106.78) \\
\end{tabular}}
\label{tb:eeg_block_cv}
\end{table}

\section{Discussion}
In this paper, we propose a novel nonparametric matrix response regression model to characterize the nonlinear association between a matrix response and scalar predictors. We introduce a low-rank nonparametric estimation procedure with a computationally efficient algorithm. The applications to the calcium imaging study and the electroencephalography study show superior performance of the proposed approach. Our method has wide applications to other modalities of imaging data as well. For example, one can model the dynamic functional connectivity matrices obtained from resting state or task-evoked fMRI time series. One can also apply our method to characterize the dynamic changes of the 2D brain surface data preprocessed from the MRI data (e.g. hippocampal morphometry surface measure studied in \citet{li2007hippocampal, yu2020beyond}). 

One of the main merits of our method is that our nonparametric  estimator has  a  closed-form  solution, and thus makes it computationally fast when analyzing large-scale imaging data. However, there are also several drawbacks. First, our method can only deal with low dimensional covariates. When the dimension of the covariates becomes high, it suffers from the curse of dimensionality, which is a long-standing issue in nonparametric smoothing methods. Second, our method is based on the Nadaraya-Watson estimator, and therefore it may also suffer from poor bias at the boundaries of the domain. To solve this problem, one may incorporate the low-rank estimation procedure into the local linear kernel method. However, the trade-off is that it significantly increases the computational burden as one can not find a closed form solution for the estimator. 

There are a number of important directions for future work. First, we model the dynamic functional connectivity as covariance matrices. One can also model them as dynamic network \citep{nie2017estimating, zhang2017finding} and consider the nonparametric regression for dynamic network. Second, one may also model the images as 2D functional data and study various models to characterize the time-dependence structures among the functional data objects \citep{zhang2015two,gao2016evolutionary}.

\section*{Acknowledgement}
The authors thank the editor, the associate editor and referees for their constructive and helpful comments on the earlier version of this paper. Kong's research was partially supported by the Natural Science and Engineering Research Council of Canada. Shen's research was partially supported by the Simons Foundation Award 512620 and the NSF Grant DMS 1509023.

\section*{Data Availability Statement}
The data that support the findings in this paper are available in \citet{eeg} at \url{https://archive.ics.uci.edu/ml/datasets/EEG+Database} and \citet{calcium} at \url{https://ajpete.com/software/}. 
 
\bibliographystyle{biom}
\bibliography{paper.bib}

\section*{Supporting Information}
Web Appendices, Tables, computational code, and a sample data set referenced in Sections \ref{method},\ref{theory}, \ref{simulation}, and \ref{data} are available with this paper at the Biometrics website on Wiley Online Library.

\end{document}



\def\spacingset#1{\renewcommand{\baselinestretch}%
{#1}\small\normalsize} \spacingset{1}


\setcounter{section}{0}

\spacingset{1.25} 

\begin{center}
    {\large\bf Supporting Information for ``Nonparametric Matrix Response Regression with Application to Brain Imaging Data Analysis''   \\ by Wei Hu, Tianyu Pan, Dehan Kong, and Weining Shen}
\end{center}

\vspace{.5in}

This file is organized as follows. Section \ref{propositionproof} contains the proof of Proposition 2. Section \ref{usefullemmas} includes useful lemmas that will be used for the proof. Section \ref{curvature} verifies the strong convexity condition of the loss function $L(\cdot)$. The proofs of theorems are included in Section \ref{theoremproof}. Sections 6 and 7 give the proof of equivalence between optimizations results (4) and (5) and Algorithm 1 in the main paper. Additional simulation result is presented in Section 8. 
  
\section{Proof of Proposition 2}\label{propositionproof} \citet{efron2004estimation} has shown a general formula for the degree of freedom as
\begin{equation*}
df = \sum_{i=1}^n\sum_{j=1}^p\sum_{k=1}^q\text{cov}(\hat Y_{ijk}, Y_{ijk})/\sigma^2,
\end{equation*}
where $Y_{ijk}$ and $\hat Y_{ijk}$ are the $(j,k)$-th element of the $Y_i $ and $\hat{Y}_i$, respectively. By Stein's theory of unbiased risk estimation \citep{stein1981estimation}, $\text{cov}(\hat Y_{ijk}, Y_{ijk}) = \sigma^2 E(\frac{\partial \hat Y_{ijk}}{\partial Y_{ijk}})$. Then an unbiased estimator of the degree of freedom is 
\begin{equation*}
\hat{df} =  \sum_{i=1}^n\sum_{j=1}^p\sum_{k=1}^q\frac{\partial \hat Y_{ijk}}{\partial Y_{ijk}} = \sum_{i=1}^n\text{tr}\left( \frac{\partial\text{vec}(\hat{Y}_i)}{\partial\text{vec}(Y_i)^T}\right).
\end{equation*}

Note that $\hat{Y}_i(\tilde\lambda)$ is a function of $\hat{Y}_{LSi}$, where $\hat{Y}_{LSi}$ is the usual least squared estimator of $Y_i$ (i.e., the NW estimator). Then we have 
\begin{equation*}
\begin{split}
\frac{\partial \text{vec}(\hat{Y}_i(\tilde\lambda))}{\partial \text{vec}(Y_i)^T} &= D(\hat{Y}_i(\tilde\lambda))(\hat{Y}_{LSi})\times D(\hat{Y}_{LSi})(Y_i)\\
& = D(\hat Y_i(\tilde\lambda))(\hat Y_{LSi}) \times \frac{K_H(X_i - X_i)}{\sum_{j=1}^n K_H(X_i - X_j)} I,
\end{split}
\end{equation*}
where $D(A)(B) =  \frac{\partial \text{vec}(A)}{\partial \text{vec}(B)^T}  $ for two matrices $A$ and $B$,  and $I$ is the identity matrix.

By taking the trace, we have:
\[
\begin{split}
\text{tr}\left\{\frac{\partial \text{vec}(\hat Y_i(\tilde\lambda))}{\partial \text{vec}(Y_i)^T}\right\}&= \text{tr}\left(D(\hat Y_i(\tilde\lambda))(\hat Y_{LSi})\right) \times \frac{K_H(0)}{\sum_{j=1}^n K_H(X_i - X_j)}\\
& = \hat{df}_i(\widetilde{\lambda}) \times \frac{K_H(0)}{\sum_{j=1}^n K_H(X_i - X_j)},
\end{split}
\]

where we define $\hat{df}_i(\widetilde{\lambda}) = \text{tr}\left\{D(\hat Y_i(\tilde\lambda))(\hat Y_{LSi})\right\}$. Further we assume that $\hat Y_{LSi}$ has distinct positive singular values $\sigma_{i1}> \sigma_{i2}> \cdots > \sigma_{ir} >0$ and $\sigma_{ik}= 0$ for $k> r$. 

By Theorem 3 in \citet{zhou2014regularized}, we have
\begin{equation}
\hat{df}_i(\widetilde{\lambda}) = \sum_{k=1}^q { 1}_{\{b_{ik}(\widetilde{\lambda})>0\}}\Big\{ 1 + \sum_{1\leq j\leq p, j\neq k, k \leq r_i} \frac{\sigma_{ik}(\sigma_{ik}-\widetilde{\lambda})}{\sigma_{ik}^2 - \sigma_{ij}^2} + \sum_{1\leq j\leq q, j\neq k, k \leq r_i} \frac{\sigma_{ik}(\sigma_{ik}-\widetilde{\lambda})}{\sigma_{ik}^2 - \sigma_{ij}^2}\Big\}.
\end{equation}
Therefore, 
\begin{equation*}
\hat{df}(\widetilde\lambda)= K_H(0)\sum_{i = 1}^n \frac{\hat{df}_i(\widetilde{\lambda})}{\sum_{j=1}^n K_H(X_i - X_j)}.
\end{equation*}  
  
\section{Useful lemmas}\label{usefullemmas}

\begin{flushleft}
We first re-state some lemmas that will be useful in the proof of the risk bound and rank consistency results. For simplicity of notation, we define
\[
L(Y;x) = \frac{1}{2n}\sum_{i=1}^n K_H(x-X_i)\|Y_i - Y\|_F^2.
\]
Note that sometimes we may write $L(Y;x)$ as $L(Y)$ for simplicity. Let $\nabla L(Y)$ be the gradient of $L(Y)$. Then the solution to (4) in the main paper can be written as
\[
\hat g(x) = \argmin\limits_{Y\in \mathbb R^{p\times q}} \{ L({Y;x}) + \lambda_n R(Y)\},
\]
where $R(\cdot)$ is a norm on $\mathbb R^{p\times q}$, and we denote $R^\ast(\cdot)$ as the dual norm of $R(\cdot)$. In our setting,  $R(\cdot)$ is the nuclear norm and $R^\ast(\cdot)$ is the spectral norm. 
\end{flushleft}
The first lemma is taken from \citet{Nega2012}. It provides useful general risk bound for high-dimensional regularized M-estimators. 
\begin{lemma}
When $\lambda_n \geq 2R^\ast(\nabla L(Y))$, any optimal solution $\hat Y_{\lambda_n}$ satisfies the bound
\begin{equation}
\|\hat Y_{\lambda_n} - Y\|^2 \leq 9\frac{\lambda_n^2}{k_L^2}\Phi^2(\overline M) + \frac{\lambda_n}{k_L}\{2\tau_L^2(Y) + 4R(Y_{M^\perp})\}
\end{equation}
\label{risk-bound}
\end{lemma}
Under our setting, $\Phi (\overline M) = \sup \limits_{u\in M\setminus\{0\}} \frac{R(u)}{\|u\|} = r$, $\tau_L(Y)= 0$, $R(Y_{M^\perp}) = 0$, and $k_L$ is a constant related to the strong convexity of the loss function that we will specify in Proposition 3. 

The next few lemmas are standard concentration bounds results for sub-Gaussian random variables and Gaussian matrices. 
\begin{lemma}
Let X be zero-mean, and supported on some interval [a, b] almost surely. Then X is sub-Gaussian with parameter at most $\sigma = b -a$.
\label{subgaussian}
\end{lemma}

\begin{lemma}\label{Hoeffding}
(Hoeffding bound) Suppose that $X_1,\ldots,X_n$ are independent and $X_i$ has mean $\mu_i$ and sub-Gaussian parameter $\sigma_i$ for $i=1,\ldots,n$. Then for every $t\geq 0$, we have
\[
P\left(\sum_{i=1}^n(X_i - \mu_i) \geq t\right) \leq \exp(-\frac{t^2}{2\sum_{i=1}^n\sigma_i^2}).
\]
\end{lemma}
\begin{lemma}\label{gordon}
(Gordon's theorem for Gaussian matrices) Let A be an $p\times q$ matrix whose entries are independent standard normal random variables. Denote $s_{\min}(A)$ and $s_{\max}(A)$   as the smallest and the largest singular value of A respectively.  Assume $p\geq q$ without loss of generality. Then
\[
\sqrt p  - \sqrt q  \leq E\{s_\text{min}(A)\} \leq  E\{s_\text{max}(A)\}\leq \sqrt p +\sqrt q.
\]
\end{lemma}

\begin{lemma}\label{concentration}
Let $Y\sim N(0,I_{d\times d})$ be a d-dimensional Gaussian random variable. Then for any function F: $\mathbb R^d \to \mathbb R$ with Lipschitz constant L, i.e. $|F(x) - F(y)| \leq L\|x - y\|$ for all $x,y \in \mathbb R^d$, we have for any $t>0$,
\[
P\left\{|F(Y) - E(F(Y))| \geq t\right\} \leq 2\exp(-\frac{t^2}{2L^2}).
\] 
\end{lemma}
The next few results are taken from \citet{bach2008consistency}. They will be used for proving rank consistency. 
\begin{lemma}\label{subdiff}
(Subdifferential) Suppose $Y = U\text{Diag}(\sigma)V^T$ is the singular value decomposition of $Y$, where $U\in \mathbb R^{p\times r}$ and  $V\in \mathbb R^{q\times r}$ have orthogonal columns and $\sigma\in \mathbb R^{r} $ with each element positive, then the subdifferential of  $\|\cdot\|_\ast$ is equal to 
\[
 \|\cdot\|_\ast(Y) = \{ UV^T + M, \text{  such that  } \|M\|_2\leq 1, U^TM = 0 \text{ and } MV = 0\}.
\]
\end{lemma}

\begin{lemma}\label{derivative}
(Directional derivative) The directional derivative at $Y = U\text{Diag}(\sigma)V^T$ is equal to:
\[
\lim_{t\to 0^+}\frac{\|Y+ t\Delta\|_\ast - \|Y\|_\ast}{t} = \text{tr} (U^T\Delta V) + \|U_\perp^T\Delta V_\perp\|_\ast.
\]
\end{lemma}

\begin{lemma}\label{lowsemi}
Assume Y has rank $r < \min\{p, q\}$ with ordered singular value decomposition $Y = U\text{Diag}(\sigma)V^T$. If $\frac{4}{s_r}\|\Delta\|_2^2 < \|(I - UU^T)\Delta(I - VV^T)\|_2$, then $\text{rank }(Y + \Delta) > r$. 
\end{lemma}

\begin{lemma}\label{optimize}
Assume $\Sigma$ is any invertible matrix and Y has singular value decomposition $Y = U\text{Diag}(\sigma)V^T$. Then the unique global minimizer of 
\[
\text{vec}(\Delta)^T\Sigma\text{vec}(\Delta) + \text{tr}U^T\Delta V + \|U_\perp^T\Delta V_\perp\|_\ast
\]
satisfies $U_\perp^T\Delta V_\perp = 0$ if and only if
\[
\text{vec}(\Lambda) = \|((V_\perp \otimes U_\perp)^T\Sigma^{-1}(V_\perp \otimes U_\perp))^{-1}((V_\perp \otimes U_\perp)^T\Sigma^{-1}(V\otimes U\text{vec}(I))\|_2 \leq 1.
\]

Moreover, the optimal solution $\Delta$ satisfies
\[
\text{vec}(\Delta) = -\Sigma^{-1}\text{vec}(UV^T - U_\perp\Lambda V_\perp^T).
\]
\end{lemma}

\section{Curvature and strong convexity}\label{curvature}
One of the major ingredients in the proof for the risk bound result is the strong convexity of the loss function. This is described using $\delta L(Y) = L(Y + \Delta) - L(Y) - <\nabla L,  \Delta>$, the remainder of the first-order Taylor expansion along some direction $\Delta$. Then we have the following proposition stating that with high probability, the loss function $L$ is strongly convex. 
\begin{proposition}
With probability of at least $1 -  \exp(-\frac{nf(x)^2h^{2s}}{32k_{\max}^2})$, 
$\delta L(Y) \geq k_L \|\Delta\|_F^2$, where $k_L = \frac{f(x)}{4} - \frac{C_{k}}{2}h^{\alpha_1}$.
\end{proposition}
\begin{proof}
To satisfy restricted strong convexity condition, we must have $\delta L(Y) \geq k_L\|\Delta\|_F^2$ for $\Delta \in \zeta(\Delta)$, where $k_L >0 $ is some constant, and $\zeta(\Delta) = \{\Delta | \|\pi_{M^\perp}(\Delta)\|_\ast \leq 3\|\pi_{\overline M}(\Delta)\|_\ast\}$, where $M^\perp$ and $\overline M$ are defined in Section 2.2 of \citet{Nega2012}. Then we have 
\[
\begin{split}
\delta L(Y) &= L(Y + \Delta) - L(Y) - <\nabla L,  \Delta>\\
=&  \frac{1}{2n}\sum_{i=1}^n K_H(x - X_i)\|Y_i - Y - \Delta\|_F^2 - \frac{1}{2n}\sum_{i=1}^n K_H(x - X_i)\|Y_i - Y\|_F^2\\
& +  \frac{1}{n}\sum_{i=1}^n K_H(x - X_i)<Y_i - Y - \Delta, \Delta>\\
=&  \frac{1}{2n}\sum_{i=1}^n K_H(x - X_i) <\Delta, \Delta>,
\end{split}
\]
where the inner product $<\cdot, \cdot>$ is defined by $<A,B> = \sum_{i,j=1}^{p,q} A_{ij}B_{ij}$. Therefore we just need to show $\frac{1}{2n}\sum_{i=1}^n K_H(x - X_i) \geq k_L$ with high probability.

By Assumption 3, $K_H(x - X_i) \in [0, h^{-s}k_{\max}]$ for any $x$ and $X_i$. Then  $K_H(x - X_i)  -  E(K_H(x - X_i)) \in [-h^{-s}k_{\max}, h^{-s}k_{\max}]$.

By Lemma \ref{subgaussian}, we have
\[
E_{X_i}(e^{\lambda(K_H(x - X_i) - E(K_H(x - X_i)))}) \leq e^{\frac{\lambda^2(k_{\max}-(-k_{\max}))^2}{2h^{2s}}} = e^{\frac{4\lambda^2k_{\max}^2}{2h^{2s}}}.
\]

Hence $K_H(x - X_i)  -  E(K_H(x - X_i))$ is sub-gaussian by definition.
By Lemma \ref{Hoeffding}, we further have, for any $t > 0$,
\[
P\left\{\frac{1}{n}\sum_{i=1}^n K_H(x - X_i) - E(K_H(x - X_i)) \leq -t\right\}\leq\exp\left(-\frac{nt^2h^{2s}}{8k_{\max}^2}\right).
\]

Meanwhile, 
\begin{equation}\label{concern1}
P\left\{\frac{1}{n}\sum_{i=1}^n K_H(x - X_i) - (E(K_H(x - X_i)) - f(x)) - f(x) \leq -t\right\}\leq\exp\left(-\frac{nt^2h^{2s}}{8k_{\max}^2}\right).
\end{equation}

By classical kernel density estimation theory, e.g., \citep{van1998asymptotic}, there exists $C_{k} \geq 0$ such that $|E(K_H(x - X_i)) - f(x)| \leq C_{k}h^{\alpha_1}$.
For simplicity, set $t = \frac{f(x)}{2}$ and $k_L = \frac{f(x)}{4} - \frac{C_{k}}{2}h^{\alpha_1}$, which is positive given $h$ is small enough and $f(x)$ is lower bounded due to Assumption 3. Therefore, with probability of at least $1 -  \exp(-\frac{nf(x)^2h^{2s}}{32k_{\max}^2})$, $\frac{1}{2n}\sum_{i=1}^n K_H(x - X_i) \geq k_L$. This completes the proof. 
\end{proof}

\section{Proof of Theorem 3}\label{theoremproof}

Denote $R^\ast(\cdot)$ as the dual norm of $R(\cdot)$. In our case, $R^\ast(\cdot)$ is the spectral norm, which is defined as the largest singular value of a matrix. We have
\[
\begin{split}
R^\ast(\nabla L(Y,x))=  &R^\ast\left(\frac{1}{n}\sum_{i=1}^n K_H(x-X_i)(Y_i - Y)\right)\\
=&R^\ast\left(\frac{1}{n}\sum_{i=1}^n K_H(x-X_i)(g(X_i) + \epsilon_i - g(x)\right)\\
\leq &R^\ast\left(\frac{1}{n}\sum_{i=1}^n K_H(x-X_i)(g(X_i) - g(x))\right) + R^\ast\left(\frac{1}{n}\sum_{i=1}^n K_H(x-X_i)\epsilon_i\right).
\end{split}
\]

Therefore, for any $t > 0$,
\begin{equation}
\begin{split}
&P\left\{\left\|\frac{1}{n}\sum_{i=1}^n K_H(x-X_i)(Y_i - Y)\right\|_2 > t\right\}\\
\leq& P\left\{ \left\|\frac{1}{n}\sum_{i=1}^n K_H(x-X_i)\epsilon_i \right\|_2 + \left\|\frac{1}{n}\sum_{i=1}^n K_H(x-X_i)(g(X_i) - g(x))  \right\|_2 > t \right\} \\
\leq&P\left\{\left\|\frac{1}{n}\sum_{i=1}^n K_H(x-X_i)\epsilon_i \right\|_2 > t/2\right\}+  P\left\{\left\|\frac{1}{n}\sum_{i=1}^n K_H(x-X_i)(g(X_i) - g(x)) \right\|_2 > t/2\right\}.\\
\label{split}
\end{split}
\end{equation}

First we look at $P\left\{\|\frac{1}{n}\sum_{i=1}^n K_H(x-X_i)\epsilon_i\|_2 > t_1\right\}$, where $t_1>0$. Note that
\begin{align*}
E\left(\left\|\frac{1}{n}\sum_{i=1}^n K_H(x-X_i)\epsilon_i \right\|_2\right) = E\left(E\left(\left\|\frac{1}{n}\sum_{i=1}^n K_H(x-X_i)\epsilon_i\right\|_2 \mid \{X_i\}_{i=1}^n\right)\right).
\end{align*}
Denote $K_H(x-X_i)$ as $c_i$. Then 
\begin{align} 
E(\|\frac{1}{n}\sum c_i\epsilon_i\|_2) &= E(\frac{\|\sum c_i\epsilon_i\|_2/n}{\sqrt{\sigma^2\sum_{i=1}^n c_i^2/n^2}}) \times \sqrt{\frac{\sigma^2 \sum_{i=1}^n c_i^2}{n^2}}\nonumber \\
&\leq \sigma \frac{\sqrt p + \sqrt q}{n} \times (\sum_{i=1}^n c_i^2)^{1/2},\label{split_prob}
\end{align}
where the last inequality is due to Lemma \ref{gordon} since entries of $\frac{\|\sum_{i=1}^n c_i\epsilon_i\|_2/n}{\sqrt{\sigma^2\sum_{i=1}^n c_i^2/n^2}} $ are i.i.d. standard normal random variables. Then
\begin{align*}
E\left(\|\frac{1}{n}\sum_{i=1}^n K_H(x-X_i)\epsilon_i\|_2\right)&\leq \sigma \frac{\sqrt p + \sqrt q}{n} E\Big((\sum_{i=1}^n K_H(x-X_i)^2)^{1/2}\Big)\\
&\leq \sigma \frac{\sqrt p + \sqrt q}{n} \Big(E(\sum_{i=1}^n K_H(x-X_i)^2)\Big)^{1/2}\\
& = \sigma \frac{\sqrt p + \sqrt q}{\sqrt n} \Big(E K_H(x-X_1)^2\Big)^{1/2}\\
& \leq 2  \sigma \frac{\sqrt p + \sqrt q}{\sqrt n}  \Big(E K_H(x-X_1)^2I(\|x- X_i\|_\infty = O(h))\Big)^{1/2}\\
& \leq 2 \sigma \frac{\sqrt p + \sqrt q}{\sqrt n} \Big(\frac{k_\text{max}^2}{h^{2s}}P(\|x- X_i\|_\infty = O(h))\Big)^{1/2}\\
&\leq 2 \sigma \frac{\sqrt p + \sqrt q}{\sqrt n} \Big(\frac{k_\text{max}^2}{h^{2s}}C_f h^s\Big)^{1/2}\\
& =  2 \sigma k_\text{max}C_f^{1/2} \frac{\sqrt p + \sqrt q}{\sqrt n h^{s/2}},
\end{align*}
where we have used the fact that $K_H(x - X_i)$ is negligible once $\| x- X_i\|_\infty \gg h $ and $P\Big(\|x - X_i\|_\infty  = O(h)\Big) = C_fh^s$ since the density function of x is bounded from above. Then we have
\begin{align}
&P\left\{\|\frac{1}{n}\sum_{i=1}^n K_H(x-X_i)\epsilon_i\|_2 - E(\|\frac{1}{n}\sum_{i=1}^n K_H(x-X_i)\epsilon_i\|_2) > t_1\right\}\nonumber \\
= & E\left(P\left\{\|\frac{1}{n}\sum_{i=1}^n K_H(x-X_i)\epsilon_i\|_2 - E(\|\frac{1}{n}\sum_{i=1}^n K_H(x-X_i)\epsilon_i\|_2) > t_1\mid \{X_i\}_{i=1}^n\right\}\right)\nonumber \\
= & E\Big(P\Big\{\frac{\|\frac{1}{n}\sum_{i=1}^n K_H(x-X_i)\epsilon_i\|_2}{(\sigma^2\sum_{i=1}^n  c_i^2)^{1/2}/n} - E\left(\frac{\|\frac{1}{n}\sum_{i=1}^n K_H(x-X_i)\epsilon_i\|_2}{(\sigma^2\sum_{i=1}^n c_i^2)^{1/2}/n}\right) \nonumber \\
& ~~~~~~~~~ > \frac{t_1}{(\sigma^2\sum_{i=1}^n c_i^2)^{1/2}/n} \mid \{X_i\}_{i=1}^n\Big\}\Big)\nonumber \\
\leq & E\left(\exp\Big(-\frac{t_1^2}{2\sigma^2\sum_{i=1}^n K_H(x-X_i)^2/n^2}\Big)\right)\label{1lip}\\
\leq& \exp\left(-\frac{t_1^2}{4\sigma^2E\Big( K_H(x-X_i)^2\Big)/n}\right)\label{slln}\\
= &  \exp\left(-\frac{nt_1^2}{4\sigma^2E\Big( K_H(x-X_i)^2 I(\|x -X_i\|_\infty = O(h))\Big)}\right) \nonumber \\
\leq  & \exp\left(-\frac{nt_1^2h^{s}}{4C_f k_\text{max}^2\sigma^2}\right).\label{term1}
\end{align}

In the above derivation, \eqref{1lip}  is due to Lemma \ref{concentration} and the fact that spectral norm is $1$-Lipschitz. Moreover, \eqref{slln} holds since $\frac{\sum_{i=1}^n K_H(x- X_i)^2}{n}$ converges to $E( K_H(x- X_i)^2)$ by strong law of large numbers and hence is bounded by 2$E( K_H(x- X_i)^2)$ with probability tending to 1 as the variance of $K_H(x-X_1)^2$ is finite.

Then we take a look at the second term in \eqref{split}. By strong law of large numbers,  continuity of $\|\cdot\|$ and continuous mapping theorem, we have
\begin{align*}
& \|\frac{1}{n}\sum_{i=1}^n K_H(x - X_i)(g(X_i) - g(x))\|_2\\
&~~ \leq 2\| E\Big(K_H(x - X_i)(g(X_i) - g(x))\Big)\|_2\\
&~~\leq \| 2E\Big(K_H(x - X_i)(g(X_i) - g(x))\Big)\|_F\\
&~~ =  \| 2E\Big(K_H(x - X_i)(g(X_i) - g(x))I(\|X_i - x\|_\infty = O(h))\Big)\|_F\\
&~~\leq  2\frac{k_\text{max}}{h^s}\| E\Big(|g(X_i) - g(x)|I(\|X_i - x\|_\infty = O(h))\Big)\|_F\\
&~~\leq  2\frac{k_\text{max}}{h^s}P\{\|X_i - x\|_\infty= O(h)\}\sqrt{pq}CMh^{\alpha_2}\\
&~~ \leq 2C_fCMk_\text{max}\sqrt{pq}h^{\alpha_2}, 
\end{align*}
where $M$ comes from the fact that $\|X_i - x\|_\infty \leq Mh$ with probability tending to 1 as $\|X_i - x\|_{\infty} = O_p(h)$. 

By matching  $2C C_f M k_\text{max}\sqrt{pq}h^{\alpha_2}$ with $2 \sigma k_\text{max}C_f^{1/2} \frac{\sqrt p + \sqrt q}{\sqrt n h^{s/2}}$, we set  $t_1 = 2C_fCMk_\text{max}\sqrt{pq}h^{\alpha_2}.$ Then

\[
\begin{split}
&P\left\{\|\frac{1}{n}\sum_{i=1}^n K_H(x-X_i)\epsilon_i\|_2 > 2CC_f M k_\text{max}\sqrt{pq}h^{\alpha_2} + \sigma k_\text{max}C_f^{1/2} \frac{\sqrt p + \sqrt q}{\sqrt n h^{s/2}}\right\}\\
\leq & P\left\{\|\frac{1}{n}\sum_{i=1}^n K_H(x-X_i)\epsilon_i\|_2 - E(\|\frac{1}{n}\sum_{i=1}^n K_H(x-X_i)\epsilon_i\|_2) > 2CC_fMk_\text{max}\sqrt{pq}h^{\alpha_2} \right\}\\
\leq  & \exp(-\frac{nt_1^2h^{s}}{4 C_fk_\text{max}^2\sigma^2})\\
= &\exp(-\frac{C_fC^2M^2pqnh^{2\alpha_2+ s}}{\sigma^2}).
\end{split}
\]

Denote $2 \sigma k_\text{max}C_f^{1/2} \frac{\sqrt p + \sqrt q}{\sqrt n h^{s/2}}$ by $t_2$. We obtain
\[
P\left\{\|\frac{1}{n}\sum_{i=1}^n K_H(x-X_i)(Y_i - Y)\|_2 > 2( t_1 + t_2)\right\}\leq \exp(-\frac{C_fC^2M^2pqnh^{2\alpha_2+ s}}{\sigma^2}).
\]

By Lemma \ref{risk-bound}, when $\lambda_n = 4( t_1 + t_2)$, we have
\[
\|\hat Y_{\lambda_n} - Y\|_F^2 \leq 9 \frac{16(t_1 + t_2)^2}{(\frac{f(x)}{4} - \frac{C_{k}}{2}h^{\alpha_1})^2}r \leq C \lambda_n^2 r,
\]
with probability of at least $1 - \exp(-\frac{C_fC^2M^2pqnh^{2\alpha_2+ s}}{\sigma^2})-  \exp(-\frac{nf(x)^2h^{2s}}{32k_{\max}^2}) $, 
where $t_1 = 2C_fCMk_\text{max}\sqrt{pq}h^{\alpha_2}$ and $t_2 = 2 \sigma k_\text{max}C_f^{1/2} \frac{\sqrt p + \sqrt q}{\sqrt n h^{s/2}}$.

\section{Rank consistency: proof of Theorem 4}
We first state and prove two useful propositions. 
\begin{proposition}
$-\frac{1}{n}\sum_{i=1}^n K_H(x- X_i)(Y_i - Y)$ and $Y$ have simultaneous singular value decompositions.
\end{proposition}
\begin{proof}
By Lemma \ref{subdiff}, the minimizer of of $L({Y;x}) + \lambda R(Y)$ satisfies
\[-\frac{1}{n}\sum_{i=1}^n K_H(x- X_i)(Y_i - Y) + \lambda(UV^T + M) = 0.\]
where $Y$ has singular value decomposition  $Y = U\text{Diag}(\sigma)V^T$(with strictly positive singular value vector $\sigma$) and $U^TN = 0$, $MV = 0$, $\|M\|_2 \leq 1$, which completes the proof.
\end{proof}

\begin{proposition}
Let $\hat g(x)$ be a global minimizer of Eq. \eqref{nuclearobjective} and assume $\hat g(x) = g(x) + \lambda_n \hat\Delta$. Then $\hat\Delta$ converges in probability to the unique global minimizer $\Delta$ of
\[
 \min_{\Delta\in \mathbb R^{p\times q}}\frac{f(x)}{2}\|\Delta\|_F^2 +  \text{tr} (U^T\Delta V) +\|U_\perp^T\Delta V_\perp\|_\ast.
\]
Moreover, we have
\begin{align*}
\hat g(x) = &g(x) + \lambda_n \Delta +   O_p\big(n^{-1/2}h^{-s/2}\lambda_n\min(p,q)\big)  + O_p\big(\lambda_n^2\min(p,q)^2\big)I \\
&+  O_p(\sqrt{pq\min(p,q)}) (n^{-1/2}h^{-s/2}+ h^{\alpha_2} ).
\end{align*}
\end{proposition}
\begin{proof}
Under the assumption $\hat g(x) = g(x) + \lambda_n \hat\Delta$, we have that
 \[
\argmin\limits_{Y\in \mathbb R^{p\times q}}\left\{\frac{1}{2n}\sum_{i=1}^n K_H(x- X_i)\|Y_i  - Y\|_F^2 + \lambda_n \|Y\|_\ast\right\}
\]
is equivalent with
\[
\begin{split}
&\argmin\limits_{\Delta\in \mathbb R^{p\times q}} \left\{\frac{1}{2n}\sum_{i=1}^n K_H(x- X_i)\|Y_i  - g(x) - \lambda_n\Delta\|_F^2 + \lambda_n \|g(x) +\lambda_n\Delta\|_\ast - \lambda_n\|(g(x)\|_\ast\right\}\\
=&\argmin\limits_{\Delta\in \mathbb R^{p\times q}} \Big\{\frac{\lambda_n^2}{2n}\sum_{i=1}^n K_H(x- X_i)\|\Delta\|_F^2 - \frac{\lambda_n}{n}\sum_{i=1}^n K_H(x-X_i) < Y_i - g(x), \Delta> \\
& \quad \quad \quad \quad \quad \quad \quad \quad +\lambda_n\|g(x) +\lambda_n (\Delta\|_\ast - \|g(x)\|_\ast) \Big\}\\
= &\argmin\limits_{\Delta\in \mathbb R^{p\times q}} \Big\{\frac{1}{2n}\sum_{i=1}^n K_H(x- X_i)\|\Delta\|_F^2 - \frac{1}{n\lambda_n}\sum_{i=1}^n K_H(x-X_i) < Y_i - g(x), \Delta> \\
& \quad \quad \quad \quad \quad \quad \quad \quad  +\frac{\|g(x) +\lambda_n\Delta\|_\ast - \|g(x)\|_\ast}{\lambda_n}\Big\}. \\
\end{split}
\]

Denote
\begin{align*}
V_n(\Delta) & =  \frac{1}{2n}\sum_{i=1}^n K_H(x- X_i)\|\Delta\|_F^2 - \frac{1}{n\lambda_n}\sum_{i=1}^n K_H(x-X_i) < Y_i - g(x), \Delta> \\
& ~~~~~~~~~+ \frac{\|g(x) +\lambda_n\Delta\|_\ast - \|g(x)\|_\ast}{\lambda_n}.
\end{align*}

First we treat the term $ \frac{1}{n\lambda_n}\sum_{i=1}^n K_H(x-X_i) < Y_i - g(x), \Delta>$. Note that
\begin{align*}
&| \frac{1}{n\lambda_n}\sum_{i=1}^n K_H(x-X_i) < Y_i - g(x), \Delta>| \\
\leq & \frac{1}{\lambda_n} \left\|\frac{1}{n} \sum_{i=1}^n K_H(x-X_i)(Y_i - g(x))\right\|_F\|\Delta\|_F\\
 \leq &\frac{1}{\lambda_n} \left\| \frac{1}{n} \sum_{i=1}^n K_H(x-X_i)(g(x_i)-g(x))\right\|_F\|\Delta\|_F + \frac{1}{\lambda_n}\| \frac{1}{n} \sum_{i=1}^n K_H(x-X_i)\epsilon_i\|_F\|\Delta\|_F \\
 \leq & \frac{2}{\lambda_n} \| E\big(K_H(x-X_i)(g(x_1)-g(x)\big)\|_F\|\Delta\|_F +  \frac{1}{\lambda_n} O_p(\frac{\sqrt{pq}} {\sqrt {nh^s}})\|\Delta\|_F \\
 \leq & \frac{2C_fCMk_{\max}\sqrt{pq}h^{\alpha_2}}{\lambda_n}\|\Delta\|_F + \frac{1}{\lambda_n}O_p(\frac{\sqrt{pq}}{\sqrt{nh^s}} )\|\Delta\|_F.
 \end{align*}

Assume $g(x)$ has singular value decomposition $g(x) = U\text{Diag}(\sigma)V^T$ with positive singular value vector $\sigma$, then by Lemma \ref{derivative},
\[
\frac{\|g(x) +\lambda_n\Delta\|_\ast - \|g(x)\|_\ast}{\lambda_n} =  \text{tr} (U^T\Delta V) + \|U_\perp^T\Delta V_\perp\|_\ast + O_p(\lambda_n \| \Delta\|_F^2).
\]
Therefore
\begin{align*}
V_n(\Delta)  = & \frac{f(x)}{2}\|\Delta\|_F^2 +   O_p(n^{-1/2}h^{-s/2})\|\Delta\|_F^2  + \text{tr} (U^T\Delta V) +\|U_\perp^T\Delta V_\perp\|_\ast  + O_p(\lambda_n \| \Delta\|_F^2)\\
&  +   2C_fCMk_{\max}\sqrt{pq}h^{\alpha_2}\lambda_n^{-1}|\Delta\|_F   + \frac{1}{\lambda_n}O_p(\frac{\sqrt{pq}}{\sqrt{nh^s}} )\|\Delta\|_F \\
= & V(\Delta) +  O_p(n^{-1/2}h^{-s/2})\|\Delta\|_F^2 + O_p(\lambda_n \| \Delta\|_F^2)+2C_fCMk_{\max}\sqrt{pq}h^{\alpha_2}\lambda_n^{-1} \|\Delta\|_F\\
&+   \frac{1}{\lambda_n}O_p(\frac{\sqrt{pq}}{\sqrt{nh^s}} )\|\Delta\|_F \\
\leq & V(\Delta) +  O_p\big(n^{-1/2}h^{-s/2}\min(p,q)\big)\|\Delta\|_2^2  + O_p(\lambda_n\min(p,q)^2 \| \Delta\|_2^2) \\
&+  2C_fCMk_{\max}\sqrt{pq}h^{\alpha_2}\min(p,q)^{1/2}\lambda_n^{-1}  \|\Delta\|_2+ \frac{1}{\lambda_n}O_p\left(\frac{\sqrt{pq\min(p,q)}}{\sqrt{nh^s}} \right)\|\Delta\|_2,
\end{align*}
where $V(\Delta) =  \frac{f(x)}{2}\|\Delta\|_F^2 +  \text{tr} (U^T\Delta V) +\|U_\perp^T\Delta V_\perp\|_\ast$.
Then we have for any $M_0 > 0$, 
\begin{align*}
E \sup_{\|\Delta\|_2 \leq M_0} & |V_n(\Delta) - V(\Delta)|  = O_p\big(n^{-1/2}h^{-s/2}\min(p,q)\big)M_0^2 +  O_p(\lambda_n\min(p,q)^2 )  M_0^2\\
 &+ \frac{2C_fCMk_{\max}\sqrt{pq}h^{\alpha_2}\min(p,q)^{1/2}}{\lambda_n} M_0  + \frac{1}{\lambda_n}O_p\left(\frac{\sqrt{pq\min(p,q)}}{\sqrt{nh^s}} \right) M_0.
\end{align*}

Suppose that $V(\Delta)$ reaches the minimum at a bounded point $\Delta_0\neq 0$. Then by Markov inequality, in the ball $\|\Delta\|_2 \leq 2 \|\Delta_0\|_2$, $V_n(\Delta)$ reaches its local minimum with probability tending to 1. Since $V_n$ is convex, the local minimum is also a global one. This completes the proof. 
\end{proof}

\noindent {\bf Proof of Theorem 4}
Let $\hat g(x)$ be a global minimizer of Eq.\eqref{nuclearobjective}. In Lemma \ref{optimize}, we can choose $\Sigma$ as  $ f(x)/2$, then 
\[
\|((V_\perp \otimes U_\perp)^Tf(x)^{-1}(V_\perp \otimes U_\perp))^{-1}((V_\perp \otimes U_\perp)^Tf(x)^{-1}(V\otimes U\text{vec}(I))\|_2  = 0.
\]
Therefore the solution of $\min V(\Delta)$ satisfies $U_\perp^T\Delta V_\perp = 0$. Moreover,  $\Delta = -2f(x)^{-1}(UV^T)$.

From previous discussion, we have $\hat g(x) = g(x) + \lambda_n\Delta + o_p(\lambda_n)$, where $\hat g(x)$ has singular value decomposition $\tilde U\text{Diag}(\tilde s) \tilde V^T$. We denote $\tilde U_0$ and $\tilde V_0$ as the singular vectors corresponding to all but the $r$ largest singular values.

Then we have
\begin{align*}
&-\tilde U_0^T\frac{1}{n}\sum_{i=1}^n K_H(x- X_i)(Y_i - \hat g(x))\tilde V_0 \\
= &-\tilde U_0^T\Big(\frac{1}{n}\sum_{i=1}^n K_H(x- X_i)(g(X_i) -g(x)) + \frac{1}{n}\sum_{i=1}^n K_H(x- X_i)\epsilon_i \\
&~~~~~~~-\frac{\lambda_n}{n}\sum_{i=1}^n K_H(x- X_i)\Delta - \frac{o_p(\lambda_n)}{n}\sum_{i=1}^n K_H(x- X_i) \Big)\tilde V_0\\
=&-\tilde U_0^T\Big( Ch^{\alpha_2} +  O_p(n^{-1/2}h^{-s/2})  - \lambda_n f(x)\Delta   - \lambda_n O_p(n^{-1/2}h^{-s/2})\Delta  \\
& ~~~~~~~~~~ -o_p(\lambda_n)O_p(n^{-1/2}h^{-s/2})\Big)\tilde V_0\\
= &~\tilde U_0^T(\lambda_n f(x)\Delta)\tilde V_0 +   o_p(\lambda_n/\sqrt{pq}),
\end{align*}
where the last equation is due to assumption that $\sqrt{pq}h^{\alpha_2}\min(p,q)^{1/2}\lambda_n^{-1} \to 0$ and  $\lambda_n^{-1} O_p\left(\frac{\sqrt{pq\min(p,q)}}{\sqrt{nh^s}} \right)\to 0$, further indicating  $h^{\alpha_2} = o(\lambda_n/\sqrt{pq})$ and  $O_p(n^{-1/2}h^{-s/2}) = o_p(1/\sqrt{pq})$ if $p \wedge q$ is finite, and $h^{\alpha_2} = o(\lambda_n/\sqrt{pq (p \wedge q)})$ and  $O_p(n^{-1/2}h^{-s/2}) = o_p(1/\sqrt{pq (p \wedge q)})$ if $p \wedge q \rightarrow \infty$. 

Since $\tilde U_0\tilde U_0^T$ and $\tilde V_0\tilde V_0^T$ converge to $U_0U_0^T$ and $V_0V_0^T$ respectively in probability, we have
\begin{align*}
\left\|\tilde U_0^T\frac{1}{n}\sum_{i=1}^n K_H(x- X_i)(Y_i - \hat g(x))\tilde V_0 \right\|_2 &= \left\|\tilde U_0\tilde U_0^T\frac{1}{n}\sum_{i=1}^n K_H(x- X_i)(Y_i - \hat g(x))\tilde V_0\tilde V_0^T \right\|_2\\
& = \lambda_n f(x)\|U_\perp U_\perp^T\Delta V_\perp V_\perp^T\|_2 + o_p(\lambda_n)\\
& = o_p(\lambda_n).
\end{align*}
Therefore $\left\|\tilde U_0^T\frac{1}{n}\sum_{i=1}^n K_H(x- X_i)(Y_i - \hat g(x))\tilde V_0 \right\|_2$ is strictly less than $\lambda_n$ with probability tending to one, which means $\text{rank} (\hat g(x)) \leq r$. Therefore we obtain rank consistency.

\section{Proof of Equivalence between (4) and (5) in the main paper}

Denote $Y_i=[Y_{i}^{(j,k)}]_{p\times q}$, and $Y=[Y^{(j,k)}]_{p\times q}$. By the definition of $\hat{g_{nw}}(x)$ in (2), we notice that finding a $Y$ which minimizes
$$\frac{1}{2}||\hat{g_{nw}}(x)-Y||_F^2+\frac{n\lambda_n}{\sum_{i=1}^nK_H(x-X_i)}||Y||_*$$
is equivalent to minimize
$$\frac{1}{2n}\sum_{i=1}^nK_H(x-X_i)\left \| \frac{\sum_{i=1}^nK_H(x-X_i)(Y_i-Y)}{\sum_{i=1}^nK_H(x-X_i)}\right\|_F^2+\lambda_n||Y||_*.$$
Then it is good enough to show that minimizing the objective function above is the same as minimizing the objective function
$$\frac{1}{2n}\sum_{i=1}^nK_H(x-X_i)||Y_i-Y||_F^2+\lambda_n||Y||_*.$$
By taking the difference between these two objective functions, we have
$$f(x,Y)=\frac{1}{2n}\sum_{i=1}^nK_H(x-X_i)||Y_i-Y||_F^2-\frac{1}{2n}\sum_{i=1}^nK_H(x-X_i)\left \|\frac{\sum_{i=1}^nK_H(x-X_i)(Y_i-Y)}{\sum_{i=1}^nK_H(x-X_i)}\right\|_F^2.$$
Now we take the partial derivative of $f(x,Y)$ with respect to $Y^{(j,k)}$ for every $j=1,2\ldots,p$, and $k=1,2\ldots,q$, and it yields 
\begin{align*} 
\frac{\partial f(x,Y)}{\partial Y^{(j,k)}} &=\frac{1}{2n}\frac{1}{\sum_{i=1}^nK_H(x-X_i)}\frac{\partial \sum_j\sum_k(\sum_{i=1}^nK_H(x-X_i)(Y^{(j,k)}-Y_i^{(j,k)}))^2}{\partial Y^{(j,k)}}\\
&~~~~-\frac{1}{2n}\sum_{i=1}^nK_H(x-X_i)\frac{\partial\sum_j\sum_k (Y^{(j,k)}-Y_i^{(j,k)})^2}{\partial Y^{(j,k)}}\\
&=\frac{1}{2n}\frac{1}{\sum_{i=1}^nK_H(x-X_i)}(2\sum_{i=1}^nK_H(x-X_i)(Y^{(j,k)}-Y_i^{(j,k)})\sum_{i=1}^nK_H(x-X_i))\\
&~~~~-\frac{1}{2n}\sum_{i=1}^n K_H(x-X_i)2(Y^{(j,k)}-Y_i^{(j,k)})\\
&=\frac{1}{n}\sum_{i=1}^nK_H(x-X_i)(Y^{(j,k)}-Y_i^{(j,k)})-\frac{1}{n}\sum_{i=1}^n K_H(x-X_i)(Y^{(j,k)}-Y_i^{(j,k)})\\
&=0.
\end{align*}
This implies that the difference between the two objective functions of interest is a constant with respect to every element of $Y$, and therefore justifies the equivalence of (4) and (5) in the main paper.

\section{Justification of using Algorithm 1 to solve optimization (4) in the main paper}
In this section, we show that Algorithm 1 does provide a valid solution to the optimization problem (4) in the main paper. First, note that optimization problem (4) is equivalent with (5) in the main paper. Second, note that optimization (5) (in the main paper) is connected to the target objective function in $D_{\tau}(Y)$. In particular, the objective function in (5) (in the main paper) is 
\begin{equation}\label{5}
\begin{split}
 \hat{g}(x)
={\rm argmin}_{Y}\left\{\frac{1}{2}\|\hat{g}_{\text{nw}}(x)-Y\|^2_F+\frac{n\lambda_n}{\sum_{i=1}^n K_H(x-X_i)}\|Y\|_*\right\}.
\end{split}
\end{equation}
and in Proposition 1 (in the main paper), we have
\begin{align}\label{Dtau}
D_\tau(Y) = \arg\min_{X} \left\{\frac{1}{2}\|Y - X\|_F^2 + \tau \|X\|_\ast\right\}.
\end{align}
It is clear that $Y$ in \eqref{Dtau} is $\hat{g}_{\text{nw}}(x)$ in \eqref{5}, $X$ in \eqref{Dtau}  is $Y$ in \eqref{5}, and the $\tau$ in \eqref{Dtau} is $\frac{n\lambda_n}{\sum_{i=1}^n K_H(x-X_i)}$. Therefore we can directly apply Proposition 1, with $Y = \hat{g}_{\text{nw}}(x)$ as the input matrix, then perform a singular value decomposition on $\hat{g}_{\text{nw}}(x)$ (Step 1 of Algorithm 1), set $\tau = \frac{n\lambda_n}{\sum_{i=1}^n K_H(x-X_i)}$ and apply the soft-thresholding operation on each non-zero singular values of $\hat{g}_{\text{nw}}(x)$ (Step 2 of Algorithm 1), and finally update the input matrix $\hat{g}_{\text{nw}}(x)$ with the new ``thresholding'' singular values (Step 3 of Algorithm 1). Based on Proposition 1 (the proof of which is given by  \citet{cai2010singular}), the output from Step 3 of Algorithm 1 does give a valid solution to optimization (4) in the main paper. 

\section{Additional simulation results}
Table 1 presents the sensitivity analysis result described in Section 5.2 of the main paper. A complete list of computational codes and data sets used in this paper is available at \url{https://github.com/vivifox/Code}.

\begin{table} 
\def~{\hphantom{0}}
\caption{Simulation results when the low rank structure is not present: mean of integrated test error and associated standard errors obtained from our
method, our method calculated with modified effective sample size (ESS-adjusted), NW estimator, and Lasso, the average selected rank and true rank are reported for three different shapes $ B $. The results are based on 100 Monte Carlo replications. For Setting I and III, ESS-adjustment is not needed since the residuals are generated to be independent.}{%
\begin{tabular}{lllccccc}  \hline
Setting I & $n$ & Shape&Our method & ESS-adjusted &NW &Lasso& Selected rank\\[5pt]
& &Cross&4493 (0.71)& &6027 (0.96)&5165 (0.82)&3.58 (0.004)\\
& 200&Square&4333 (0.72)& NA &6110 (0.99)&4886 (0.84) &1.99 (0.000)\\
& &Tshape&4422 (0.76)& &6097 (0.98)&5209 (0.93)&3.18 (0.005)\\
&&&&&&&\\
&& Cross&4340 (0.48)& &5011 (0.52)&4571 (0.48) &3.99 (0.000)\\
& 500 &Square&4225 (0.41)& NA &4804 (0.48)& 4446 (0.46)&2.07 (0.003)\\
& &Tshape&4291 (0.45)& &5044 (0.51)&4588 (0.49) &3.48 (0.004)\\ \hline
Setting II & $n$ & Shape&Our method& ESS-adjusted &NW &Lasso& Selected rank  \\[5pt]
 & &Cross&4692 (2.48)& 4668 (2.21) &5974 (3.34)& 5549 (3.50)&6.63 (0.040)\\
&   200&Square&4499 (2.37)& 4480 (2.29) &5904 (3.40)&5193 (6.44)&4.41 (0.030)\\
& &Tshape&4700 (2.5)& 4677 (2.46) &6019 (3.66)&5608 (3.74)&6.65 (0.009)\\
& &&&&&&\\
& & Cross&4436 (1.23)& 4423 (1.22) &5034 (1.76)& 4784 (1.72) &7.16 (0.008)\\
&   500 &Square&4331 (1.12)& 4322 (1.08) &5037 (1.64)& 4701 (1.69)&4.99 (0.008)\\
&  &Tshape &4437 (1.23)& 4425 (1.23)  &5058 (1.60)& 4810 (1.89)&7.01 (0.008)\\ \hline
 Setting III & $n$ &Shape&Our method& ESS-adjusted &NW &Lasso& Selected rank  \\[5pt]
&  &Cross&5958 (0.99)& &7610 (1.23)& 5859 (1.01)&6.07 (0.009)\\
 &  200&Square&5502 (0.93)& NA &6723 (1.08)& 5345 (1.05)&2.70 (0.007)\\
 & &Tshape&5796 (1.04)& &7142 (1.11)& 5586 (1.14)&4.60 (0.007)\\
 &&&&&&&\\
 & &Cross&5491 (0.57)& &5565 (0.63)& 5159 (0.52) &7.30 (0.008)\\
& 500 &Square&5357 (0.56)& NA &5508 (0.58)& 4657 (0.45) &2.48 (0.006)\\
&& Tshape&5508 (0.61)& &5623 (0.59)& 4912 (0.48) &7.31 (0.007)\\ \hline
Setting IV & $n$ &Shape&Our method & ESS-adjusted &NW &Lasso& Selected rank  \\[5pt]
 & &Cross&5872 (3.73)& 5822 (3.10) &7169 (3.90)& 6066 (6.69)&5.95 (0.030)\\
& 200&Square&5595 (2.70)& 5454 (2.56) &6861 (3.57)& 5381 (3.99) &3.33 (0.011)\\
&  &Tshape&5903 (3.02)& 5864 (2.96) &7288 (3.91)& 5935 (4.39) &5.73 (0.012)\\
 &&&&&&&\\
 & &Cross&5548 (1.70)& 5481 (2.12) &5690 (2.38)& 5303 (1.98)&9.17 (0.009)\\
& 500 &Square&5334 (1.82)& 5283 (1.50) &5464 (2.12)& 4941 (2.90) &4.05 (0.016)\\
&& Tshape& 5560 (1.91)& 5489 (1.59) & 5747 (2.09)& 5320 (1.88)&9.13 (0.029)\\ \hline
 \end{tabular}} 
\end{table}

\newpage

\bibliographystyle{biom}
\bibliography{paper-ref.bib}